\documentclass[11pt,letterpaper,nofootinbib]{revtex4-2}
\usepackage[top=100pt,bottom=0pt,letterpaper]{geometry}
\usepackage{adjustbox}
\usepackage{natbib}
\usepackage[T1]{fontenc}
\usepackage{lmodern}
\usepackage[utf8]{inputenc}
\usepackage{microtype}
\usepackage{xcolor}
\usepackage{setspace}
\usepackage{float}
\restylefloat{table}

\usepackage{dcolumn}
\usepackage{lipsum}
\usepackage{graphicx,epsfig}
\usepackage{psfrag}
\usepackage{bm}
\usepackage{epstopdf}
\usepackage{latexsym}
\usepackage{multirow}
\usepackage{amsmath,amssymb}
\usepackage{enumerate}
\usepackage{hyperref}
\usepackage[FIGTOPCAP]{subfigure}
\usepackage{empheq,mathtools}

\usepackage{booktabs}
\usepackage{tabularx}
\usepackage{longtable}
\def\be {\begin{equation}}
\def\ee {\end{equation}}
\def\ba {\begin{eqnarray}}
\def\ea {\end{eqnarray}}
\def\nn {\nonumber}
\def\bc {\begin{center}}
\def\ec {\end{center}}
\newcommand{\bdm}{\begin{displaymath}}
\newcommand{\edm}{\end{displaymath}}

\def\o  {\omega}
\def\O  {\Omega}

\def\r  {\rho}
\def\th {\theta}

\def\t  {\tau}

\def\mc {\mathcal}
\def\nn {\nonumber}
\def\ra {\rightarrow}

\def\la {\label}
\def\le {\left}
\def\ri {\right}
\def\pa {\partial}
\def\f {\frac}
\def\sq {\sqrt}

\def\bi {\begin{itemize}}
\def\ei {\end{itemize}}

\def\> {\rangle}
\def\< {\langle}

\def\bc {\begin{center}}
\def\ec {\end{center}}

\usepackage{amsfonts}

\begin{document}
\title{Signature quasinormal modes of Ellis-Bronnikov wormhole embedded in warped braneworld background}

\author{Antariksha Mitra} \email[email: ]{antarikshamitra@prl.res.in}

\author{Suman Ghosh} \email[email: ]{suman.ghosh@bitmesra.ac.in}
\affiliation{Department of Physics, Birla Institute of Technology, Ranchi - 835215, India}

\begin{abstract}
We examine the quasi normal modes of Ellis-Bronnikov wormholes embedded in a warped five dimensional braneworld background and compare with it's four dimensional counterpart.
These scalar quasi normal frequencies are obtained using the WKB formula, Prony method and the direct integration method.
The signature of the warped extra dimension shows up as two distinct quasi normal ringing era, characterised by two distinct dominant quasi normal modes. Features of the latter region are similar to that observed earlier for massive scalar field in black hole background particularly the existence of arbitrarily long lived quasi normal modes.
We also discuss the how steepness of the neck of the wormhole effects the quasi normal frequencies.

\end{abstract}
\maketitle






\section{Introduction}
\label{intro}

Wormholes are known to be solutions of Einstein field equations, like the Schwarzschild black hole in vacuum, that essentially connects two distinct spacetime points within our Universe (intra-Universe) or two `parallel universes’ (inter-universe) creating a short-cut that allows `apparently faster than light' travel. 
Detailed historical account of theoretical discovery/construction  of wormholes could be found, for example, in \cite{Visser:1995cc,Alcubierre:2017pqm}. 
The original wormhole solutions were found to be {\em non-traversable} \cite{Kruskal:1959vx,Fuller:1962zza,Eardley:1974zz,Wald:1980nm}  or unstable under perturbation.
Violation of the (averaged) null energy condition is required to prevent the wormhole `throat' from collapsing and making it {\em traversable}.
This could be realised by introducing exotic matter around the throat \cite{Morris:1988cz,Lobo:2005us}. 
It appears as if such matters may have a quantum origin, but the  standard model matter seems to be inadequate for the generation of macroscopic wormholes \cite{Witten:2019qhl}.
Remarkably, plethora of wormhole constructions under the so-called `modified theories of gravity' exist in the literature that avoids the use of exotic matter \cite{Hochberg:1990is,Bhawal:1992sz,Agnese:1995kd,Fukutaka:1989zb,Ghoroku:1992tz,Furey:2004rq,Bronnikov:2009az,Lobo:2008zu,Kanti:2011jz,Kanti:2011yv,Zubair:2017oir,Shaikh:2016dpl,Ovgun:2018xys,Canate:2019spb,Myrzakulov:2015kda,Dai_2020,Rahman_2023}.
It has also been suggested that violation can be restricted to arbitrarily small region \cite{Visser:2003yf}.

The four dimensional Ellis-Bronnikov spacetimes (4D-EB) \cite{Ellis:1973yv, Bronnikov:1973fh} that employs phantom scalar field (a field with negative kinetic term) and is one of the most researched wormhole geometries since its introduction in 1973. 
Several studies on this class of model can be found in the literature, including geometry of spinning 4D-EB spacetime \cite{Chew:2016epf}, generalized spinning of 4D-EB wormhole in scalar-tensor theory \cite{Chew:2018vjp}, hairy Ellis wormholes solutions \cite{Chew:2020svi}, Ellis wormholes in anti-de Sitter space \cite{Blazquez-Salcedo:2020nsa}, stability analysis of 4D-EB solution in higher dimensional spacetime \cite{Torii:2013xba} as such.
Kar et al. presented a generalised version of 4D-EB spacetime (4D-GEB) \cite{Kar:1995jz}, where the need for exotic matter is {\em partially} evaded by introducing a new wormhole parameter, $n \geq 2$ ($n = 2$ corresponds to 4D-EB geometry). Quasi-normal modes (QNM), echoes and some other aspects of 4D-GEB wormhole are analysed in \cite{DuttaRoy:2019hij}. 

Wormholes are yet contemplated as conjectural. However, recent developments in black hole observation \cite{Vagnozzi:2022moj} have increased the possibility to distinguish a black hole from a so-called black hole mimicker such as a wormhole. 
In fact we are far from identifying a black hole from what we have observed yet \cite{Murk:2022dkt}.
In principle, one may identify wormholes through lensing effects, shadows, Einstein rings, and other phenomena \cite{Abe:2010ap, Toki:2011zu, Takahashi:2013jqa, Cramer:1994qj, Perlick:2003vg, Tsukamoto:2012xs, Bambi:2013nla} which may in turn favour modified gravity theories over general relativity.
QNMs \cite{Vishveshwara:1970zz, Macedo:2016wgh, Konoplya:2011qq, Cardoso:2003pj, Kokkotas:1999bd, Pani:2013pma,Bronnikov:2021liv,Biswas:2022wah} are one such signature that characterises-- e.g. the late time response (`ringing') of a black hole (or wormhole) under perturbation.
Dominant quasi-normal frequencies (QNFs) can be seen in the gravitational wave signals from black holes (or similar compact objects) at late times. 
They have been observed recently by LIGO/VIRGO collaborations \cite{LIGOScientific:2016aoc, LIGOScientific:2016sjg, LIGOScientific:2017bnn, LIGOScientific:2017vwq}.
Remarkably observation of multi-mode quasi-normal spectrum has been reported in \cite{Capano:2021etf}.
This allows one to determine the individual black hole/wormhole parameters involved.
Determination of QNFs with high accuracy is an important challenge and can constrain various modified gravitational theories also test strong gravity regime.

One class of the modified theories of gravity involve extra spatial dimension(s). In fundamental physics, the emergence of an additional spatial dimension is ubiquitous -- Kaluza and Klein \cite{Kaluza:1921tu, Klein:1926tv} first demonstrated it in an effort to combine gravity and electromagnetic theories for a five-dimensional (5D) gravity model in 1921 and 1926, respectively. 
Be it the string theory \cite{Green:1987sp} or in the context of symmetries of particle physics (the octonionic hypotheses) \cite{Furey:2015yxg, Baez:2001dm, Baez:2010ye, Furey:2018yyy, Furey:2018drh, Gillard:2019ygk}, the extra  dimensions seems to appear {\em naturally}. 
String theory also motivated the brane-world scenarios-- where our 4 dimensional (4D) Universe (3-brane) is embedded in a higher dimensional bulk. 
The so-called DGP models produce infra-red modification with extra dimensional gravity dominating at low energy scale \cite{Dvali:2000hr}.
Perhaps, the most popular of these models are the `warped braneworld' models \cite{Rubakov:1983bb, Gogberashvili:1998iu, Gogberashvili:1998vx,Randall:1999ee, Randall:1999vf} that generate ultra-violet modification to general relativity with extra dimensional gravity dominating at high energy scale and address the Hierarchy issue in the fundamental scales of physics.
These models, feature a non-factorisable curved 5D space-time where the 4D metric is a function of the additional dimension through a warping factor. 

Attempts to build wormhole models in higher-dimensional spacetime has began to appear recently \citep{Lobo:2007qi, deLeon:2009pu, Wong:2011pt, Kar:2015lma, Banerjee:2019ssy, Wang:2017cnd}. 
Kar \cite{Kar:2022omn} has proposed a 5D warped wormhole model where the warping chosen is largely inspired by the non-static Witten bubble.
Recently, in \cite{Sharma:2021kqb}, an EB spacetime embedded (with a decaying warp factor) in 5D warped  bulk (5D-WEB) is constructed that is supported by on-brane positive energy density matter. Though the weak energy condition is violated, the degree of violation could be made arbitrarily small. 
We further analysed the timelike trajectories and the geodesic congruences in these spacetimes in detail in \cite{Sharma:2022tiv, Sharma:2022dbx}. 
The warping factor, we assume, is that of the well-known thick brane model \cite{Dzhunushaliev:2009va, Koley:2004at, Zhang:2007ii, Ghosh:2008vc}, which is a smooth function of the extra dimension (thus there are no derivative jump or delta functions in the curvature and connections).

In this work, we determine the QNFs (using multiple techniques/algorithms) for both the 4D-(G)EB and 5D-WEB spacetimes and contrast them to distinguish the effects or signatures of the wormhole parameters and the warped extra dimension.
The following is a breakdown of the content of this article. 
In Section (\ref{sec:model}), we briefly introduce the novel 5D-W(G)EB wormhole geometry alongside it's 4D counterpart. 
In Section (\ref{sec:field-eqn}), the field equation for (scalar) perturbation of the geometry and corresponding effective potentials are derived.  
In section (\ref{sec:QNM}), we discuss various methods to solve the master equation in order to determine the time domain profile of the perturbation and QNFs. 
In Section (\ref{sec:Results}), we report the results and compare 4D and 5D models to distinguish the signature of the warped extra dimension and the wormhole parameters. 
Remarkably, we found two distinct QNM era with two different dominant QNFs. 
Finally, in Section (\ref{sec:Dis}) we summarise the work done and key results.

\section{4D-GEB and 5D-W(G)EB Spacetime}\label{sec:model}

4D-EB wormhole is a spacetime geometry constructed in presence of a phantom matter field-- one whose action contains a negative kinetic energy term. This solution is a spherically symmetric, static and geodetically complete, horizonless manifold that has a `throat' (which becomes apparent in an embedding diagram \cite{Morris:1988cz}) linking two asymptotically flat regions and is given by the following line element,
\begin{equation}
    ds^2=-dt^2+\frac{dr^2}{1-\frac{b_0^2}{r^2}}+r^2d\theta^2 +r^2\sin^2\th d\phi^2 . \la{eq:EB-1}
\end{equation}
Here $b_0$ is the wormhole's throat radius. 
The EB spacetime metric, can also be written as,
\begin{equation}
 ds^2=-dt^2+dl^2+r^2(l)~d\theta^2 +r^2(l)\sin^2\th~ d\phi^2 \la{eq:EB-2}
\end{equation}
\begin{equation}
 \mbox{with}~~   r^2(l)=l^2+b_0^2 \la{eq:r-EB}
\end{equation}
and $l$ is called the `tortoise coordinate' or `proper radial distance'.
A generalisation of the EB model (GEB) is proposed in \cite{Kar:1995jz} (which is consistent with Morris-Thorne conditions essential for a Lorentzian wormhole), given by
\begin{equation}
      ds^2=-dt^2+\frac{dr^2}{1-\frac{b(r)}{r}}+r^2 d\th^2 +r^2\sin^2\th ~d\phi^2
\end{equation}
\begin{equation}
\mbox{with}~~ b(r)=r-r^{(3-2n)}(r^n-b_0^n)^{(2-\frac{2}{n})}.
\end{equation}
The parameter $n$ takes only even values so that $r(l)$ is smooth over the complete range of $-\infty<l<\infty$. For $n=2$, we get the original EB geometry back.
The GEB metric looks much simpler in terms the tortoise coordinate
\begin{equation}
dl^2=\frac{dr^2}{1-\frac{b(r)}{r}} \implies   r(l)=(l^n+b_0^n)^{\frac{1}{n}}. \la{eq:r-GEB}
\end{equation}
Note that at the wormhole throat ($l=0$) the sole non-vanishing derivative is the $n^{th}$-order derivative of $r(l)$. The effective potential (elaborated later on) also has a non-zero $n^{th}$-derivative at $l=0$, which gives a negative value for EB model ($n=2$ case), while for all other $n$ values it provides a positive value.


The 5D warped Ellis-Bronnikov model, introduced in \cite{Sharma:2021kqb} is
\begin{equation}
ds^{2} =  e^{2f(y)} \Big[ - dt^{2} +  dl^{2} + r^{2}(l)~\big(  d\theta^{2} + \sin^{2}\th ~d\phi^{2} \big) \Big] + dy^{2}. \label{eq:5d-WGEB}
\end{equation}
In this model, $y$ is an extra dimension ($ - \infty \leq y \leq \infty$), $f(y)$ is a warp factor and the term in square bracket is the GEB space-time.
We assume, $f(y) = \pm \log[\cosh(y/y_0)]$, \footnote{for all numerical calculation we have chosen $y_0 = 1$.} which represent known thick brane solutions in presence of bulk matter fields \cite{Dzhunushaliev:2009va}. This choice also avoids jump or delta function in connections and Riemann tensors. 
In \cite{Sharma:2021kqb}, we showed that for this class of models two of the (on-brane) weak energy conditions, $\r>0$ and $\r + p >0$ is satisfied in presence of a decaying warp factor.
Further, instead of having $n>2$  in 4D-GEB, having an warped extra dimension as in 5D-WGEB model, removes the negative energy density matter completely from the 3-brane located at $y=0$ \cite{Sharma:2022dbx, Sharma:2022tiv}. 
The weak energy condition violation comes from radial pressure $\t$ as,
\be
\r(l,y) + \t(l,y) = -2 e^{-2f(y)}\f{r''(l)}{r} < 0, \la{eq:WEC-2}
\ee
 is negative everywhere. This term can be made arbitrarily small in the $b \ra 0$ limit and one may then assume quantum effects to justify such violation.
Note that, for a decaying warp factor, the pre-factor has it's minimum value at the location of the brane only.


\section{Field equation and effective potential}\la{sec:field-eqn}

The perturbations or fluctuations in a black hole or wormhole geometry may be caused by merger or gravitational interactions with other astrophysical objects or even the so-called test objects that may represent an spaceship passing through.
The scalar frequencies of these perturbation evolve via a massless Klein-Gordon equation, given by, 
\begin{equation}
  \nabla_\mu \nabla^\mu \Psi=\frac{-1}{\sqrt{-g}} \pa_\mu \left(g^{\mu\nu} \sqrt{-g} ~\pa_\nu\Psi \right) = 0, \la{eq:KG}
 \end{equation}
where $\Psi$ is the scalar (field) perturbation and $g$ is the determinant of the metric tensor involved. 
For massive scalar field perturbation with mass $m$, the right hand side of Eq. (\ref{eq:KG}) will carry a term $m^2 \psi$. 
This equation does allow (with appropriate boundary conditions) solutions having complex frequencies. 
These QNFs have a natural interpretation as gravitational radiation where the black hole/wormhole is treated as an open system.
The QNFs, by definition, are associated with specific boundary conditions which says they are purely outgoing waves at spatial infinities. 
The real part of a QNF denotes the oscillation while the  imaginary part implies damping of the field over time. 
The vector and tensor perturbations (wherever applicable) also follow a similar field equation as scalar frequencies as such.
These QNFs are also key to test stability of a wormhole geometry under perturbation. They certainly depend on the various wormholes parameters involved, thus could have distinct features in comparison with black hole as such. 
Analysis of the effective potential and determination of QNF's for the 4D-GEB model is briefly addressed in \cite{DuttaRoy:2019hij}. Below we reproduce and extend their result of the 4D scenario and then compare them with the corresponding results derived for 5D-WEB spacetime.

\subsection{4D scenario}

Since the wormhole geometry is static and spherically symmetric, one may use the following separation of variable for the field $\Psi$, in the 4D-GEB scenario, as
\begin{equation}
 \Psi (t,r,\theta,\phi)= \mc{Y}(\theta,\phi)\frac{R(r)e^{-i\omega t}}{r}, \la{eq:sep-var-4D}
\end{equation}
where $\mathcal{Y}(\theta,\phi)$ are the spherical harmonics.
This leads to a form similar to Schr$\"{o}$dinger equation in the tortoise coordinate $l$,
\begin{equation}
    \omega^2+\frac{1}{R}\frac{\pa^2R}{\pa l^2}-V_{eff}=0. \la{eq:Sch}
\end{equation}
The `effective potential' $V_{eff}$ is given by
\begin{equation}
    V_{eff}=\left[\frac{(n-1)b_0^nl^{n-2}}{(l^n+b_0^n)^2}+\frac{m(m+1)}{(l^n+b_0^n)^{2/n}}\right].  \la{eq:Veff-GEB}
\end{equation}
where, $m$ represents the azimuthal angular momentum. 
In terms of the radial coordinate, the effective potential is simply,
\begin{equation}
  V_{eff}=\left[\frac{r''}{r}+\frac{m(m+1)}{r^2}\right].
\end{equation}
Before going into the solutions of the field equation and determination of the QNFs, let us analyse the effective potential corresponding to perturbations in 4D and 5D models.


\begin{figure}
    \centering
    \includegraphics[width=0.4\textwidth, height=0.35\textwidth]{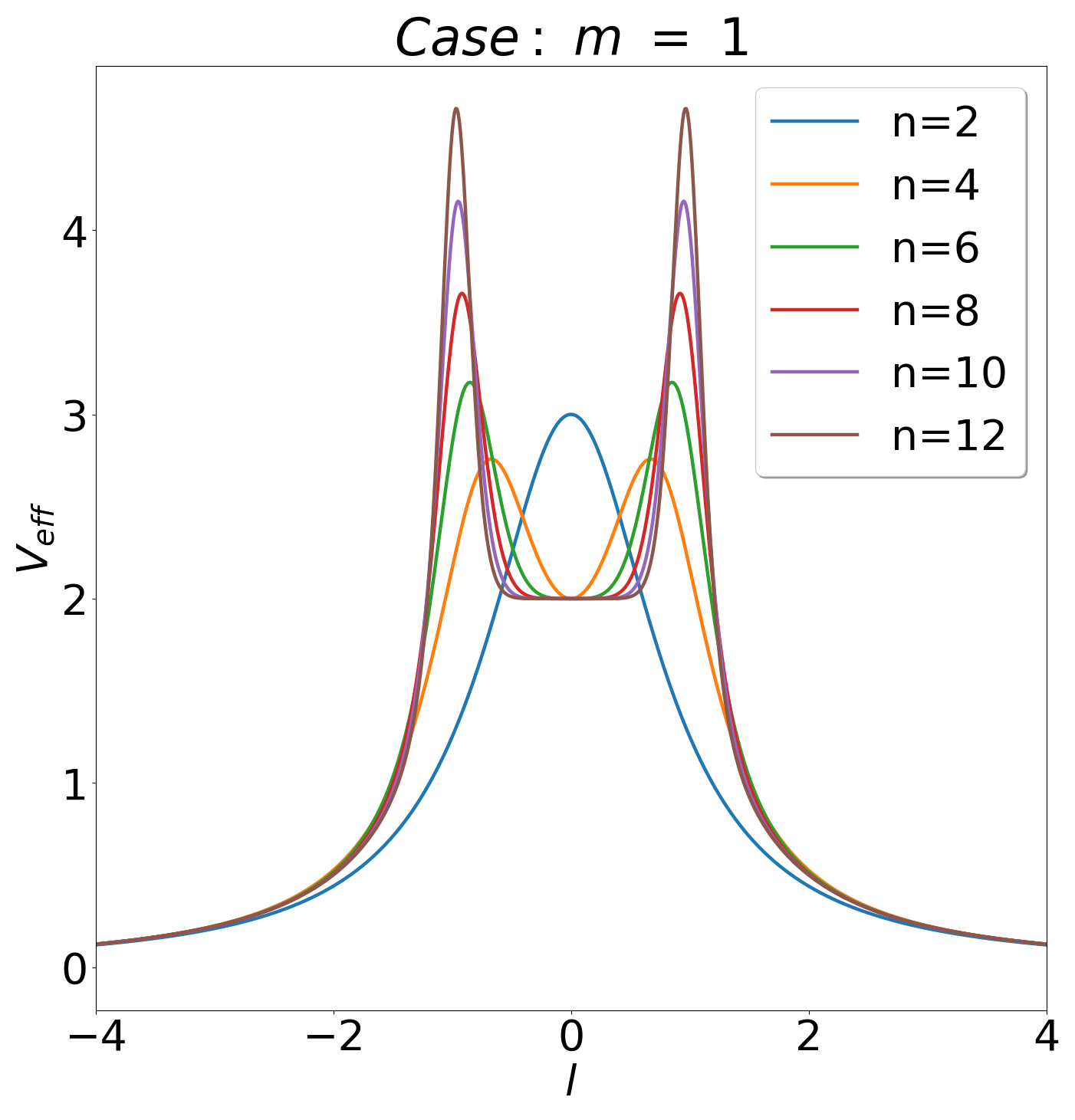}\hspace{1cm}\includegraphics[width=0.4\textwidth, height=0.35\textwidth]{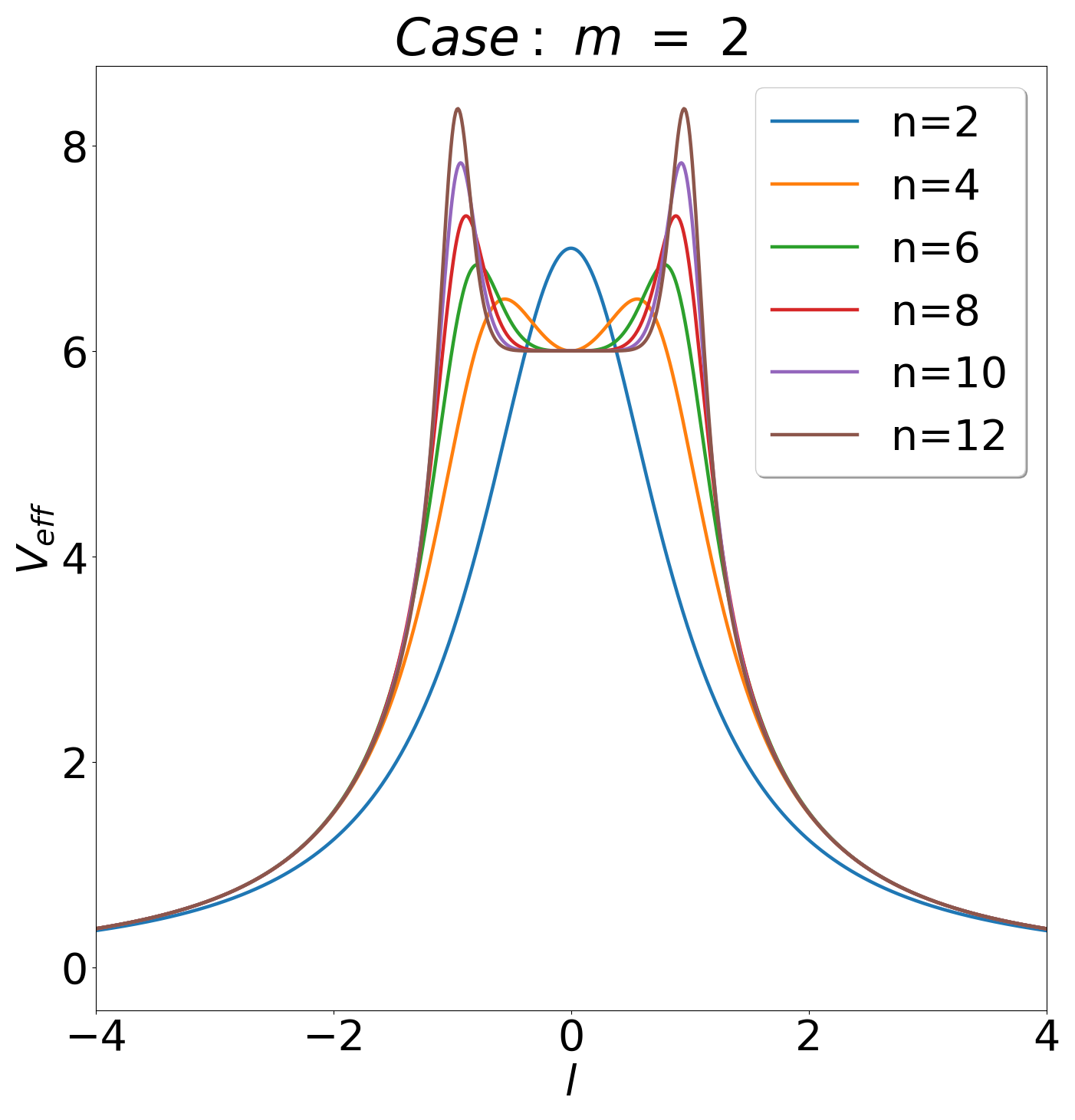}
    \includegraphics[width=0.4\textwidth, height=0.35\textwidth]{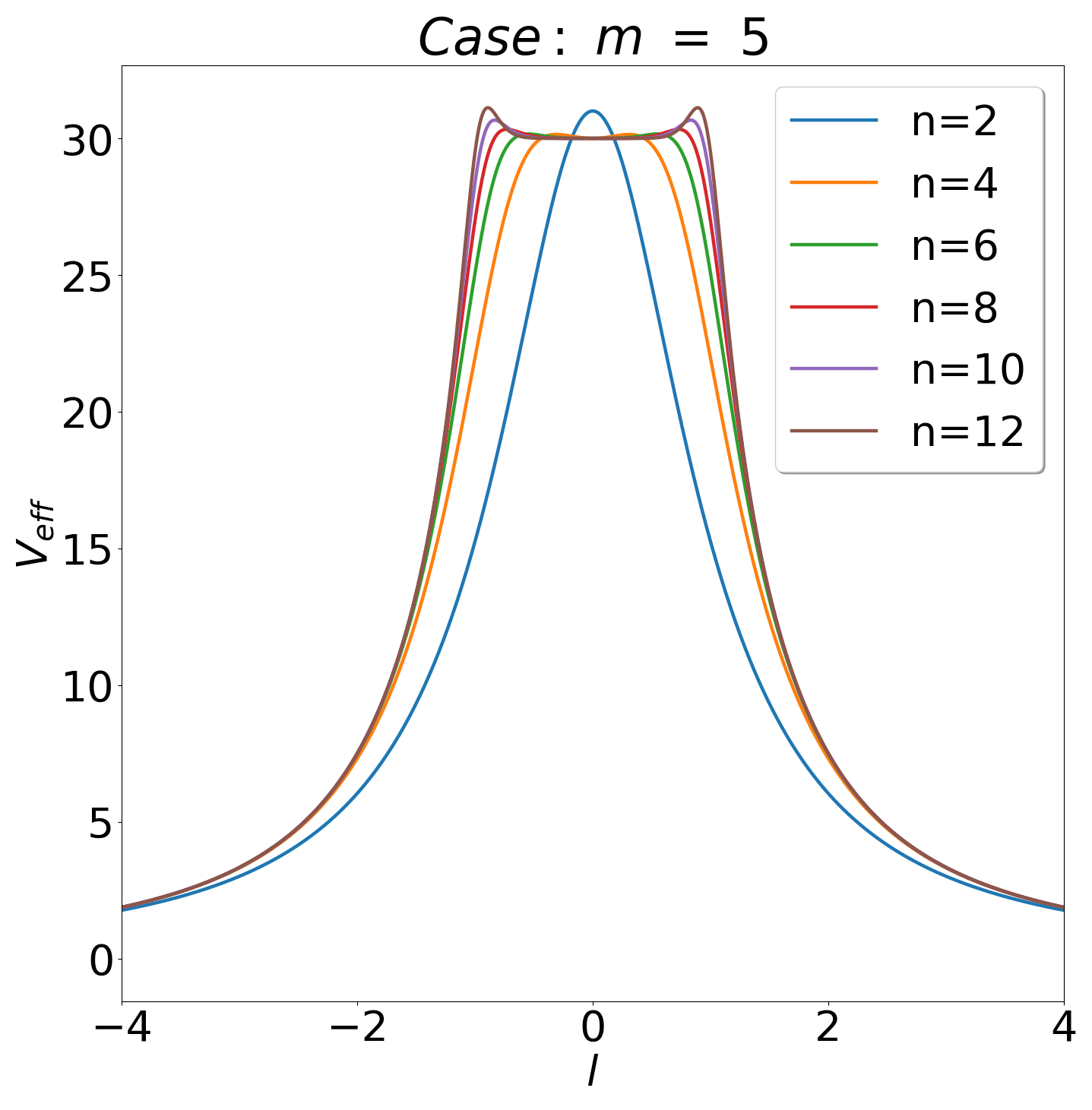}\hspace{1cm}\includegraphics[width=0.4\textwidth, height=0.35\textwidth]{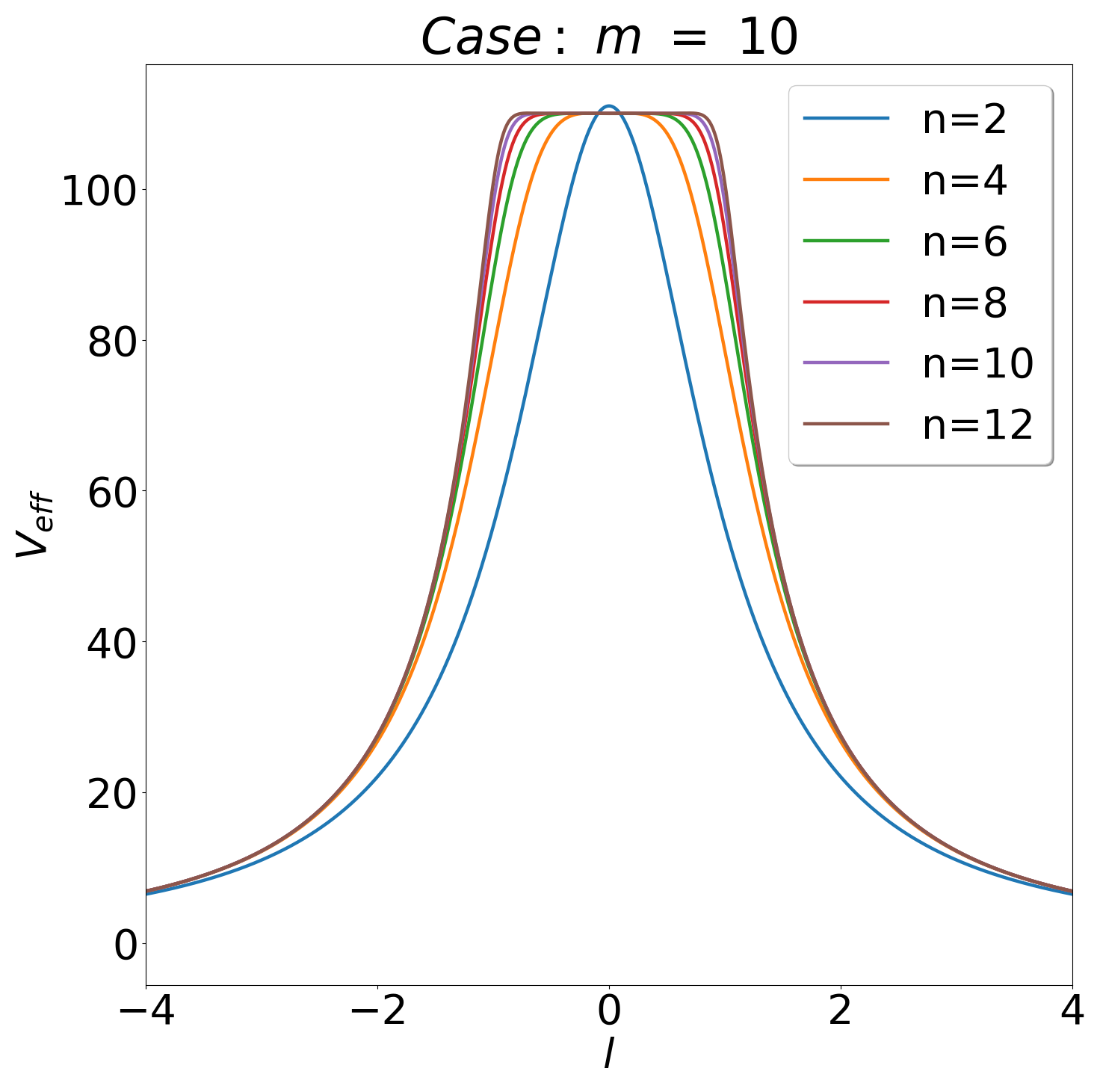} 
    \caption{Plot of effective Potential for fixed $m$ and varying $n$, (top left) $m=1$, (top right) $m=2$, (bottom left) $m=5$, (bottom right) $m=10$}
    \label{fig:Veff-4D-m-const}
\end{figure}

\begin{figure}
    \centering
    \includegraphics[width=0.4\textwidth,height  =0.35\textwidth]{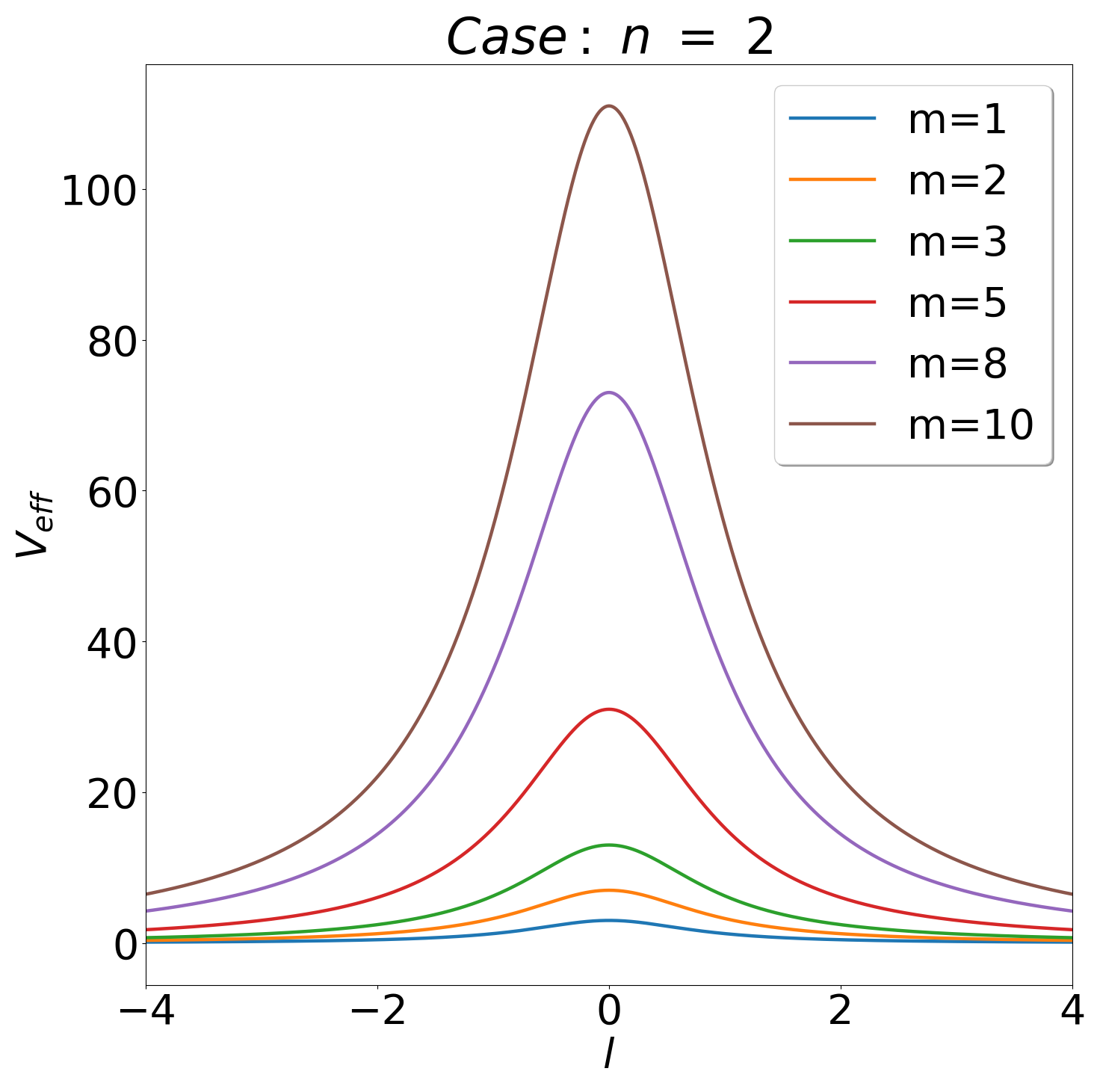}\hspace{1cm}
    \includegraphics[width=0.4\textwidth,height =0.35\textwidth]{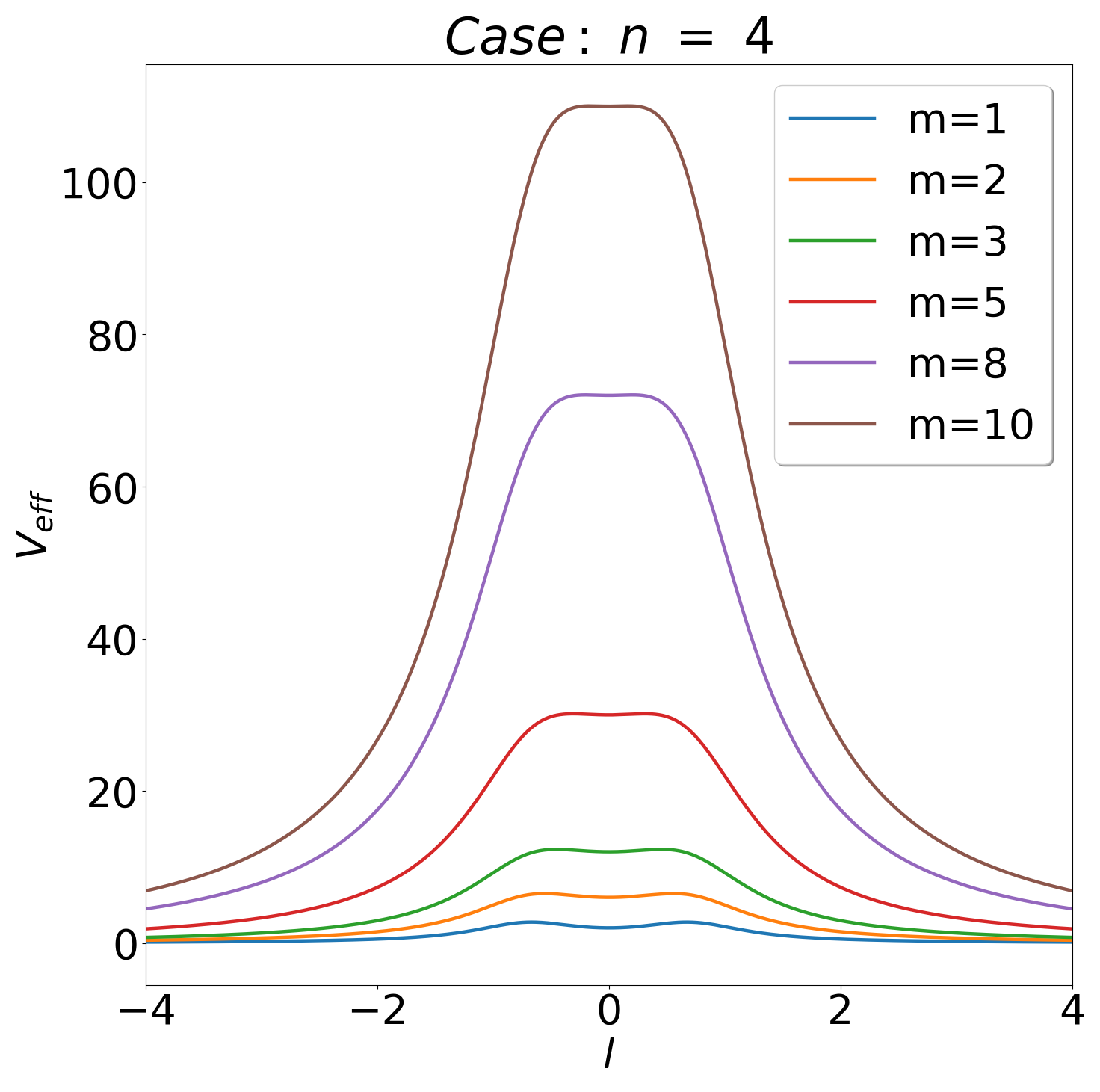} 
    \caption{Plot of effective Potential for fixed $n$ and varying $m$, (left) $n=2$, (right) $n=4$.}
    \label{fig:Veff-4D-n-const}
\end{figure}

Fig. \ref{fig:Veff-4D-m-const} shows the variation of the effective potential vs $l$ for the various 4D-GEB models (varying $n$) for four different angular frequencies $m = 1, 2,, 5, 10$. 
While the plots in Fig. \ref{fig:Veff-4D-n-const}, show the variation of effective potential vs $l$ for $n=2$ (EB case) and $n=4$ (for various $m$ frequencies).
\footnote{The throat radius $b_0$ is taken as unity for numerical evaluation.} 
Few prominent features observed from these plots are as follows.
\begin{itemize}
    \item $V_{eff}$ exhibits a single barrier for $n = 2$, while a twin barrier exists for all $n > 2$.
This particular feature, in fact, corresponds to removal of the exotic matter from the throat region.
    \item For higher $m$ frequencies, the potential increases and for $n>2$, twin-peaks merge to create a plateau-shaped single barrier. In other words, the twin barrier feature is only visible for low values of $m$. This could have important implications for the stability of the 4D-GEB model which may- be addressed elsewhere. 
    \item It is known that WKB method to determine QNFs are not suitable for twin barrier potentials. However, the potential profiles suggest that WKB formula could be useful even for $n>2$ for high $m$.
    \item All potentials vanish asymptotically which imply trivial boundary condition for QNMs.
\end{itemize}

\subsection{5D scenario}

In the 5D-WGEB spacetime given by Eq. \ref{eq:5d-WGEB}, we use the following separation of variables 
\begin{equation}
    \Psi^{5D}=Y(\theta ,\phi )e^{-i\omega t }\frac{R(r)}{r}F(y)e^{-f(y)} , \la{eq:sep_5D}
\end{equation}
where $F(y)$ and $f(y)$ are  functions only depending on $y$.
Thus the Klein-Gordon equation in 5D leads to, 
\ba
   && \left(\omega^2-\left[\frac{(n-1)b_0^nl^{n-2}}{(l^n+b_0^n)^2}+\frac{m(m+1)}{(l^n + b_0^n)^{2/n}}\right]+\frac{1}{R}\frac{\pa^2R}{\pa l^2}\right) \nn\\ 
    &&~~ = - \left[\frac{1}{F}\frac{\pa^2 F}{\pa y^2} + 2\frac{1}{F}\frac{\pa F}{\pa y}\frac{\pa f}{\pa y}-\frac{\pa^2 f}{\pa y^2}-3\left(\frac{\pa f}{\pa y}\right)^2\right]  e^{2f(y)} \la{eq:KG-5D-1}
\ea
Now taking
\begin{equation}
  -  e^{2f(y)}\left[\frac{1}{F}\frac{\pa^2 F}{\pa y^2} + 2\frac{1}{F}\frac{\pa F}{\pa y}\frac{\pa f}{\pa y}-\frac{\pa^2 f}{\pa y^2}-3\left(\frac{\pa f}{\pa y}\right)^2\right] = q^2 ,\la{eq:y}
\end{equation} 
Eq. (\ref{eq:KG-5D-1}) reduces to the form of Eq. (\ref{eq:Sch}) with the \textit{effective potential} being,
\begin{equation}
    V_{eff}=\left[\frac{(n-1)b_0^nl^{n-2}}{(l^n+b_0^n)^2}+\frac{m(m+1)}{(l^n+b_0^n)^{2/n}}\right] + q^2 \la{eq:5D-pot} .
\end{equation}

\begin{figure}[h]
    \centering
\includegraphics[width=0.5\textwidth]{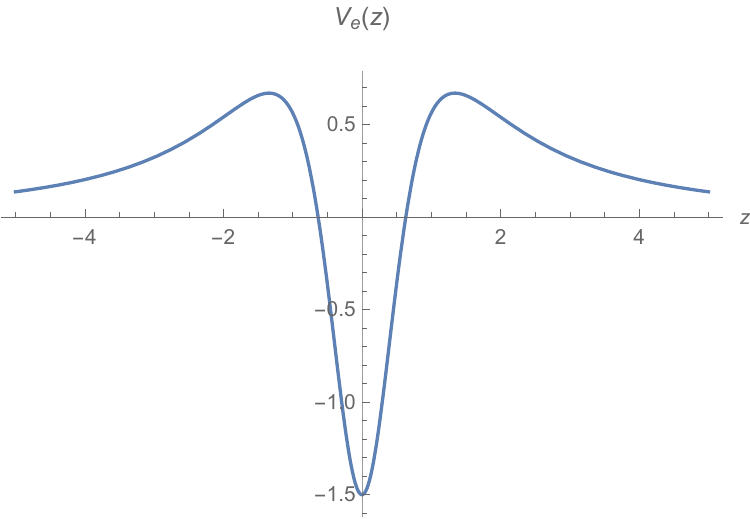} 
    \caption{Plot of potential $V_e(z)$ vs $z$.}
    \label{fig:pot_z}
\end{figure}

To find the eigenvalues $q^2$ by solving Eq. (\ref{eq:y}), let us first do a coordinate transformation given by $dz = dy ~e^{-f}$ i.e. $z = \sinh y$ (for the decaying warp factor), that leads to
\begin{equation}
    \frac{\pa^2 F}{\pa z^2} + \frac{\pa f}{\pa z} \frac{\pa F}{\pa z} -\le[\frac{\pa^2 f}{\pa z^2} + 2\left(\frac{\pa f}{\pa z}\right)^2\ri]F = -q^2 F . \la{eq:z}
\end{equation} 
Then, we use the ansatz $F(z) = G(z)\exp(-f/2)$, leading to the following simpler form
\begin{equation}
-\frac{\pa^2 G}{\pa z^2} + V_{e}(z) G = q^2 G, ~~~~ \mbox{where} ~~~~ V_{e}(z) = \f{3}{2}\frac{\pa^2 f}{\pa z^2} + \f{9}{4}\left(\frac{\pa f}{\pa z}\right)^2 = \f{3(5z^2 - 2)}{4(z^2 + 1)^2}.  \la{eq:z_Sch}
\end{equation} 
The potential $V_e(z)$ is plotted in Fig. \ref{fig:pot_z}. This potential vanishes as $z \ra \pm \infty$ which implies that positive (or real $q$) eigenvalues are a continuum. This analysis is consistent with numerical solution found using MATHEMATICA. 
Apparently, it may seem that for negative eigenvalues a discrete spectrum exists. This can be investigated with the following approximation.
The series expansion of the potential about $z=0$ is given by 
\be
V_e(z) = - \f{3}{2} + \f{27}{4}z^2 - 12z^4 + \f{69}{4}z^6 - O(z)^8 \la{eq:V_expan}
\ee
Since, for the negative part of the potential-- $|z|< 1$, we choose to ignore the terms of $O(z)^4$ onwards. This leaves us with a harmonic oscillator potential whose eigenvalues are given by
\be
E_{h.o.} = \le(n + \f{1}{2}\ri)\sq{27} - 3/2, ~~~~ n = 0,1,2... \la{eq:eigen_HO}
\ee
The ground state eigenvalue, for $n=0$, is positive in spite of the factor $-3/2$. This observation remains unchanged if we include higher order terms and find the eigenvalue numerically. Thus there are no negative eigenvalues or bound states of Eq. (\ref{eq:z_Sch}). 
\footnote{Negative eigenvalues would imply imaginary $q$-values.}
Thus the $q^2$-term effectively contributes as a effective mass in the Schr\"{o}dinger equation. This is a well-known feature of massless 5D field equation when projected on 4D as such. 
Thus the potential given by Eq. \ref{eq:5D-pot} is equivalent to the potential of massive scalar field with mass $q$ in the 4D-GEB (or pure EB) background.
However, no such studies could be found in the literature which makes it difficult compare our result with previous result.
One can say that Eq. \ref{eq:5D-pot} is unique and results in unique signature QNMs as we are going to see in what follows.
Some what similar effective potential appears in presence of a massive scalar field in a 4D black hole background \cite{Konoplya:2004wg,Zhidenko:2009zx}. 
There authors have shown that for some values of the black hole mass and the scalar field mass, purely real QNM frequencies or the so-called quasi-resonances exist.\footnote{Naturally, there is a debate whether these frequencies can be called QNF at all.}
We shall see similar results below. Horowitz and Hubeny \cite{Horowitz:1999jd} has addressed a similar problem in the sense that we also have an asymptotically non-vanishing potential (Fig. \ref{fig:pot-5d} below).
Note that the effect of the extra dimension is encoded in the eigenvalues $q^2$. If $q^2$ takes continuous values then the information about the functional form the warp factor would not be imprinted on the QNFs to be determined below.


In what follows, we shall choose suitable values of $q$ to be put in Eq. (\ref{eq:5D-pot}) and Eq. (\ref{eq:Sch}) for numerical evaluation and graphical presentations. 
Note that, $q$ has a dimension of inverse length. Therefore, its exact numerical value is less important for our purpose. So, one can set $y_0=1$ without loosing any generality. However, `$q = b_0^{-1}$' is expected to have a physical significance as we will see later.

\begin{figure}[h]
    \centering
    \includegraphics[width=0.4\textwidth]{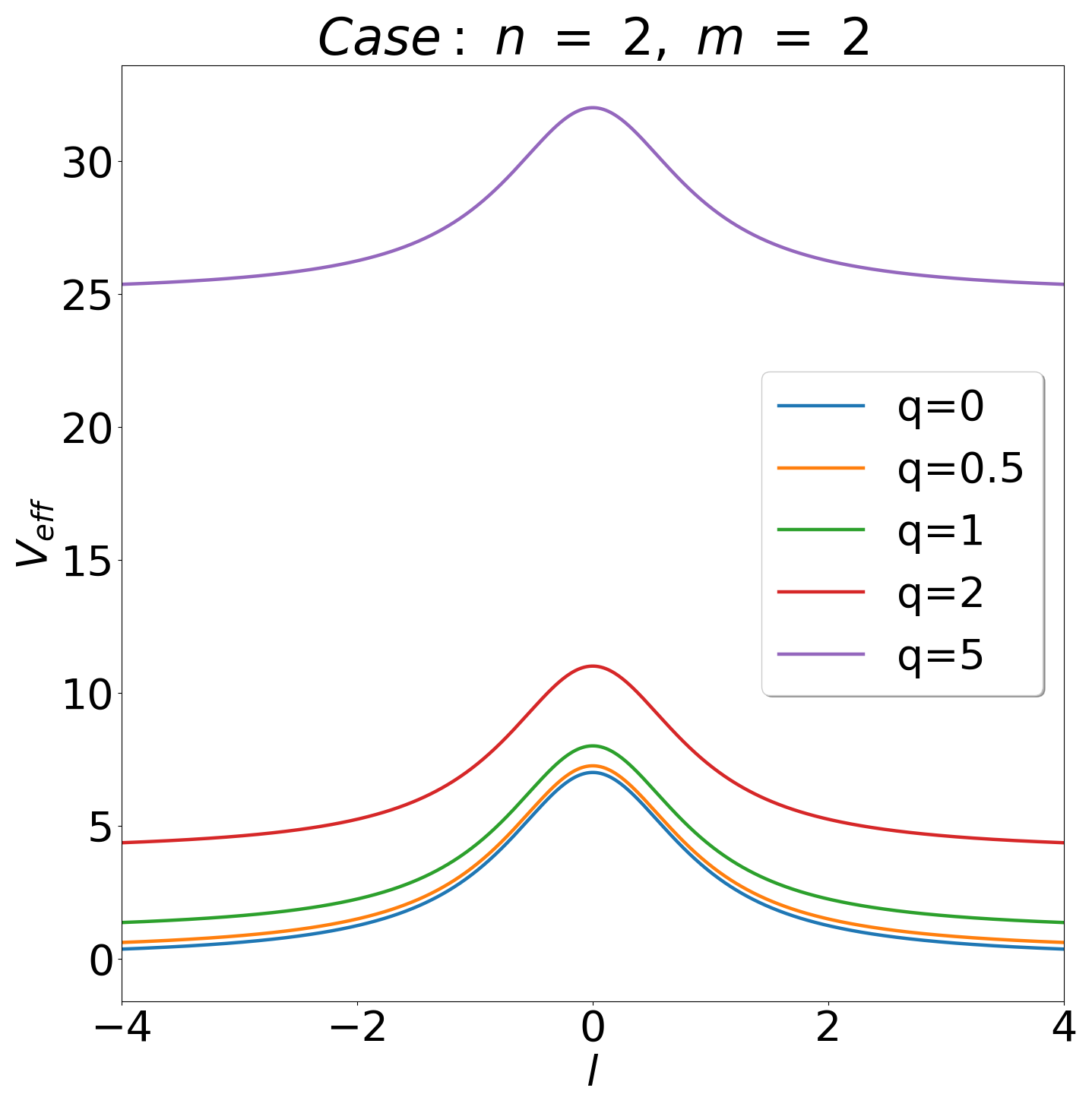}\hspace{1cm}\includegraphics[width=0.4\textwidth]{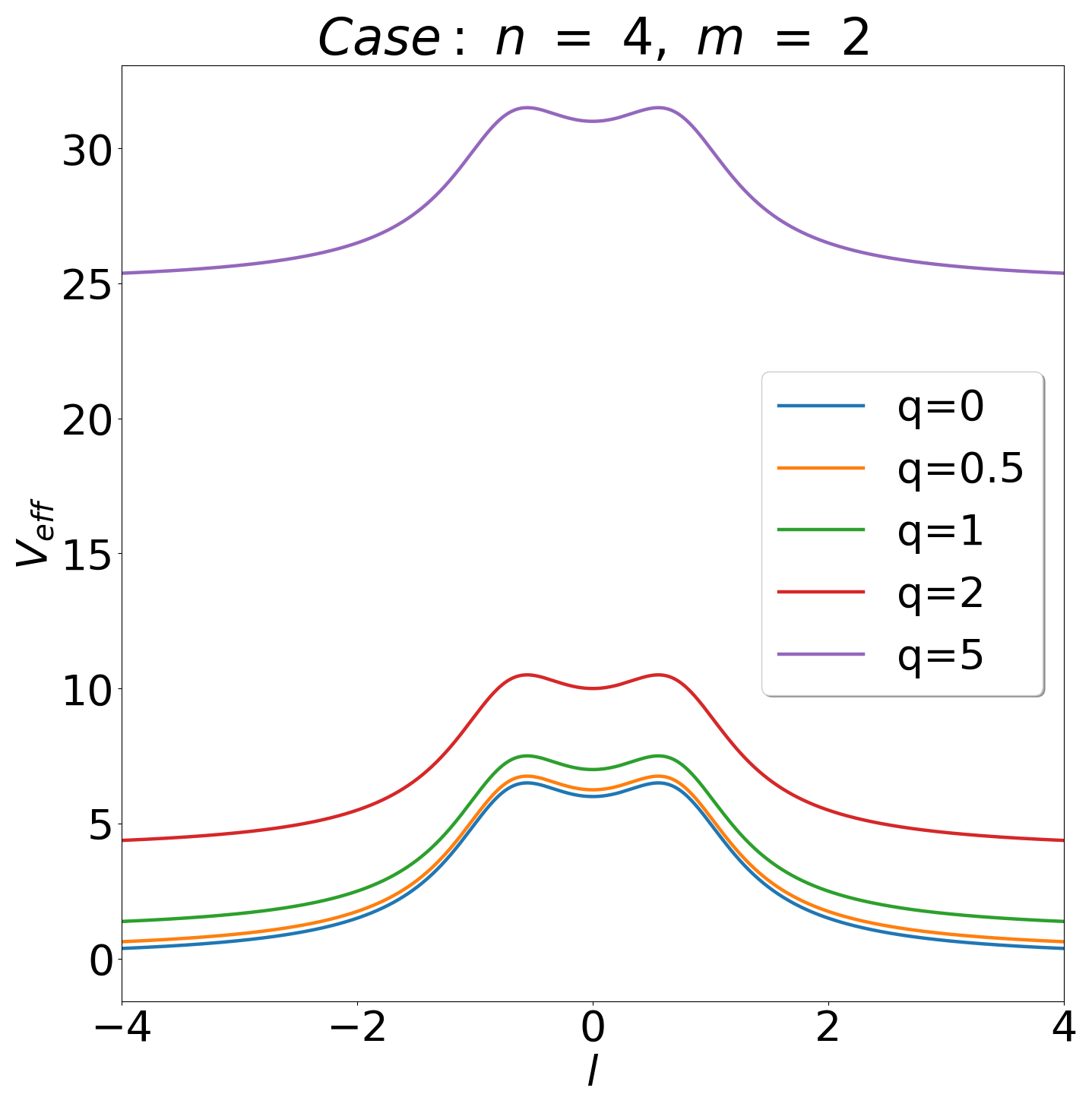} 
    \caption{Plot of effective Potential for constant $m$ and $n$ at different dimensions(varying q),(left)$ n=2,m=2$;(right) $n=4,m=2$.}
    \label{fig:pot-5d}
\end{figure}
Fig. \ref{fig:pot-5d}, shows the effective potential profile vs $l$, for various values of $q$ with fixed azimuthal angular momentum ($m=2$) for $n=2$ (WEB) and $n=4$ (a WGEB) geometries.
Due to the presence of the extra dimension is that as $l \ra \infty$, the potential does not vanish and essentially becomes equal to $q^2$ consistent with Eq. (\ref{eq:5D-pot}).
This will be reflected on the choice of boundary conditions to determine the QNFs using various methods.

\section{Time Domain Spectrum $\&$ Quasi Normal Frequencies} \la{sec:QNM}

WGEB models with a decaying warp factor does satisfy the enegy conditions even for $n=2$ or the original EB spacetime \cite{Sharma:2021kqb} where there is a single barrier only.
Which in  turn suggests that one may use WKB method to find approximate QNF values not only for 4D-EB wormholes but also for 5D-WEB ($n=2$) wormholes as well. We employ numerical methods to supplement the WKB approach also for higher accuracy in cases where the WKB approach is less efficient. 

\subsection{WKB approach}
The semi-analytical WKB approximation to derive QNFs was developed by Schutz and Will \cite{Schutz:1985km}. The method is based on matching of the asymptotic WKB solutions at spatial infinities and at the neck of the wormhole (event horizon in case of a black hole) with the Taylor expansion near the top of the potential barrier through the two turning points.
The QNF frequencies found by taking the WKB solutions upto the eikonal limit are given by the following formula \cite{Iyer:1986np,Konoplya:2003ii}
\be
w^2_p = V_0 -i \le(p+ \f{1}{2}\ri)\sq{-2V_0''} \la{eq:w_wkb}
\ee
where $V_0$ and $V_0''$ denote the values of the effective potential and its second derivative at the maximum. 
$p$ denotes the overtone number with $p = 0$ being the fundamental mode.
For our model Eq. (\ref{eq:w_wkb}) implies (for $n=2$ scenario)
\be
w^2_p = \f{m^2+m+1}{b_0^2} + q^2 - 2 i (2p+ 1) \f{\sq{m^2+m+2}}{b_0^2} \la{eq:w_wkb-1}
\ee
It is straightforward to show that for large effective mass `$q>> m$', we get purely real $w_p \sim q$ or the so-called quasi resonances which were reported for wormholes in \cite{Churilova:2019qph}. 
Numerical results will show more details. 
Here, we focus only on the fundamental frequencies and compared the WKB values with the numerical results (derived in the next section) in the tables given below. 
For a recent comprehensive review on WKB methods one may refer to \cite{Konoplya:2011qq}.
Let us now discuss the numerical methods to compute the QNFs for 4D and 5D geometries.

\subsection{Numerical approaches}
 
QNFs are complex frequencies that characterise damped oscillation of gravitational perturbations in the metric. 
There are many methods developed to determine these frequencies. Few numerical approaches are designed to find QNFs with any desired accuracy (see \cite{Konoplya:2011qq} for a review of methods), which are based on convergent procedures.
Each has it's own advantage and disadvantages. 
Developing efficient method to compute QNFs is an active area of research. 
The analytic methods, e.g. WKB method, are less accurate compared to numerical methods, for example in presence of multiple barriers (e.g. $n>2$ in GEB models) \cite{DuttaRoy:2019hij}.
The time dependent wave equation, integrated over the angular coordinates, has following form,
\begin{equation}
    V_{eff}\Psi_m(t,l)+\frac{\pa^2\Psi_m(t,l)}{\pa t^2}-\frac{\pa^2\Psi_m(t,l)}{\pa l^2}=0 \la{eq:KG-num}
\end{equation}
 
\begin{figure}[h]
    \includegraphics[width=.45\textwidth]{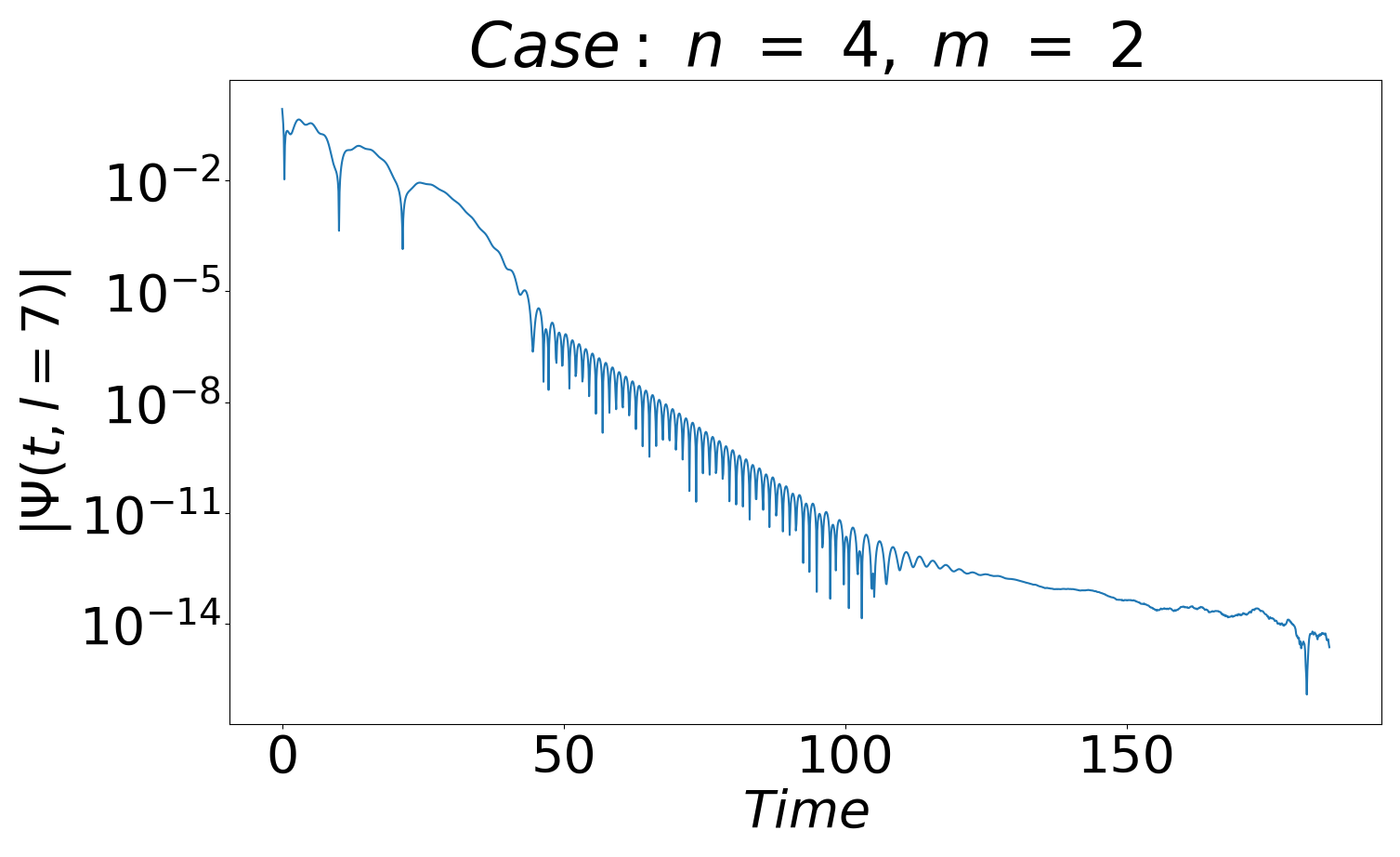} 
\hspace{.5cm}
    \includegraphics[width=.5\textwidth]{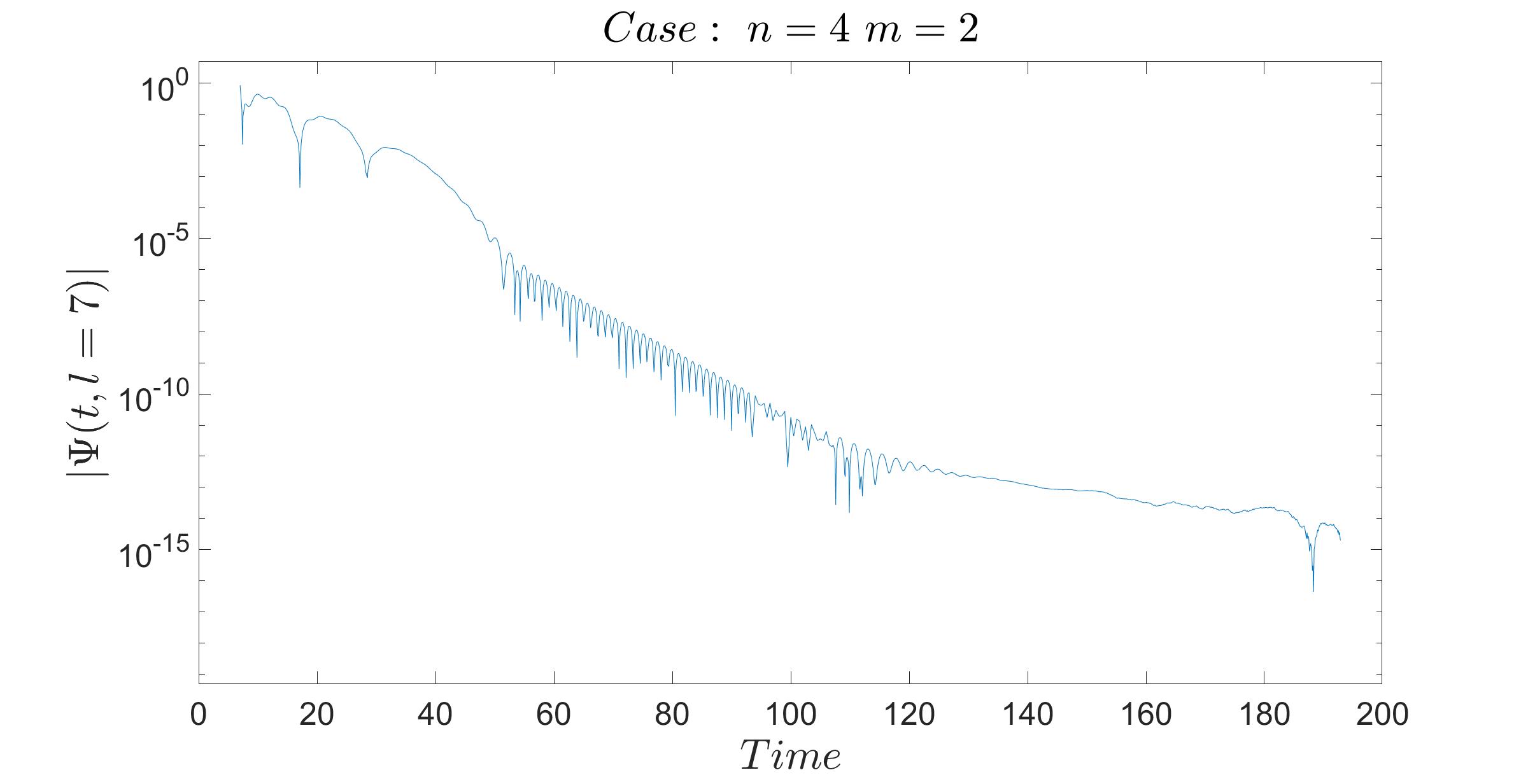} 
    \caption{Time domain Spectrum for $n=4, m=2$, using Python (left) and MATLAB (right).}
   \label{fig:Py-Mat} 
\end{figure}

In the first method, we determine the time evolution of the scalar perturbation by numerically integrating Eq. (\ref{eq:KG-num}) using the methodology presented in \cite{Konoplya:2011qq, Gundlach:1993tp}.
The essential steps are as follows.
One first adopts the light cone coordinates, $du=dt+dl$ and $dv=dt-dl$, which implies,
 \begin{equation}
    \left(4\frac{\pa^2}{\pa u\pa v}+V_{eff}(u,v)\right)\Psi_m(u,v)=0 \la{eq:KG_uv}
\end{equation}
The time evolution operator, using simple two-variable Taylor expansion reads as,
\begin{equation}
\begin{split}
  &  \exp\left(h\frac{\pa}{\pa t}\right)= \exp\left(h\frac{\pa}{\pa u}+h\frac{\pa}{\pa v}\right)\\ & = -1 + \exp\left(h\frac{\pa}{\pa u}\right)+ \exp\left(h\frac{\pa}{\pa v}\right)+\frac{h^2}{2}\left[\exp\left(h\frac{\pa}{\pa u}\right)+ \exp\left(h\frac{\pa}{\pa v}\right)\right]\frac{\pa^2}{\pa u\pa v}+\mathcal{O}(h^4)
    \end{split}  \la{eq:op_uv}
\end{equation} 
Where $h$ is the step size. Thereafter, we numerically integrate over $du$ and $dv$, ideally in the range $[0,\infty]$.  We have computed the field amplitude in the region $0 \leq u,v \leq 200$ with a step size $h = 0.01$. 
The initial condition is taken as a Gaussian distribution along $v=0$,  $\Psi(u,0)= e^{\frac{-(u-10)^2}{100}}$, and as a constant along $u=0$, $\Psi(0,v)=1/e$, such that they equate at $\Psi(0,0)$. 
This computation is performed using both Python and MATLAB to cross-check for accuracy. A particular case, with $n=4,m=2$ in the 4D-GEB model, is shown in the (log-linear) plots in Fig. \ref{fig:Py-Mat}. This shows that the efficiency of the two computing platforms are comparable. The presence of quasi-normal frequencies is clearly evident from these time domain evolution spectrum. 

From the log-linear plots in Fig. \ref{fig:Py-Mat}, one may identify three distinct stages in the time domain spectrum the \textit{initial region}, approximately for tenure $t=0-50s$ (which depends on the initial condition), the second stage, the region of our interest-- the exponential dampening, roughly during $t=50-110s$ followed by the third stage of `tail'  \cite{Konoplya:2011qq,Gundlach:1993tp}.
Note that with increase in $b_0$ value the duration quasi-normal ringing increases which suggests an decrease in the value of the QNF. This feature is also present in the 5D scenario.
In fact this can be seen by straigntforward evaluation of the WKB formula Eq. (\ref{eq:w_wkb}).
From the damped region one can extract the QNF values by the {\em Prony fitting} method (discussed below).
We have also used the {\em direct integration} method to determine the QNFs. 
Below, we briefly discuss these methodologies followed by the results tabulated. 

\subsection{Determination QNFs}\label{sec:methods}


\subsubsection{Prony Method}

\begin{figure}[h]
    \centering
    \includegraphics[width=.6\textwidth]{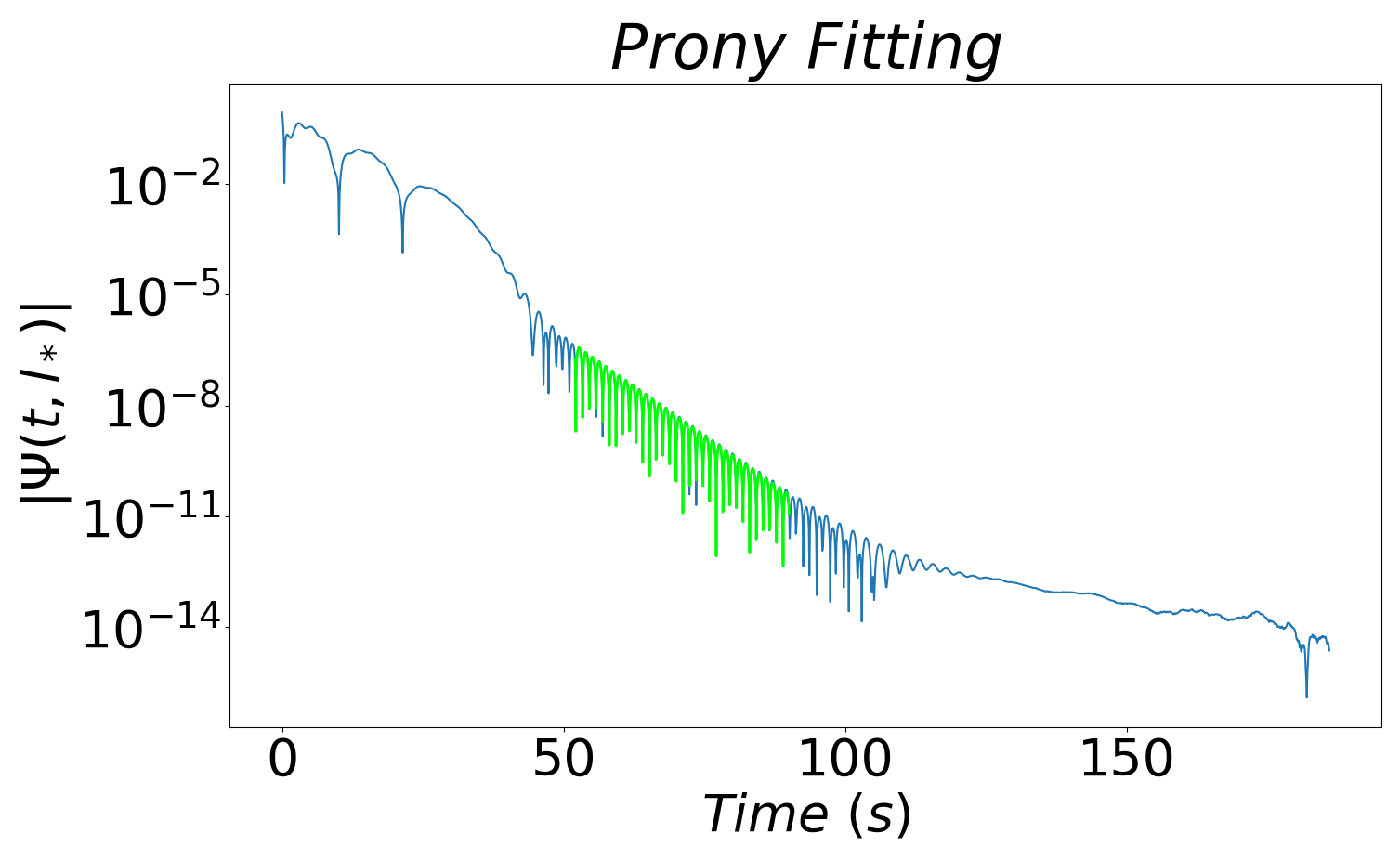} 
    \caption{Time domain Spectrum fit with dominant $\omega_{QNF}$ for n=4, m=2 using Python. }
    \label{fig:pfit-n4-m2-q0}
\end{figure}

In the prony method, the time domain profile is fitted by the function
\begin{equation}
    f(t)=\sum_{n=0}^{\infty}A_ne^{\alpha_nt}\cos{(\beta_nt)}, \la{eq:prony}
\end{equation} 
where the QNFs are given by, $\omega_{QNF}=\alpha\pm i\beta$. This technique is similar to the Fourier method but is also valid for complex frequencies and was first developed by G.R.D. Prony in 1795 \cite{Konoplya:2011qq}. There is another variation of this technique \cite{Kokkotas:1999bd}, where the function is equated with the time domain spectrum and converted into a matrix form whose roots (eigenvalues) are found to be the QNFs. This technique is used in all fields having any damped oscillatory signal processing.
We used both of these approaches for reliability (using both Python and MATLAB). 
An example of the Prony fitting is depicted in the Fig. \ref{fig:pfit-n4-m2-q0}. 

\begin{figure}[h]
    \centering
    \includegraphics[width=0.45\textwidth]{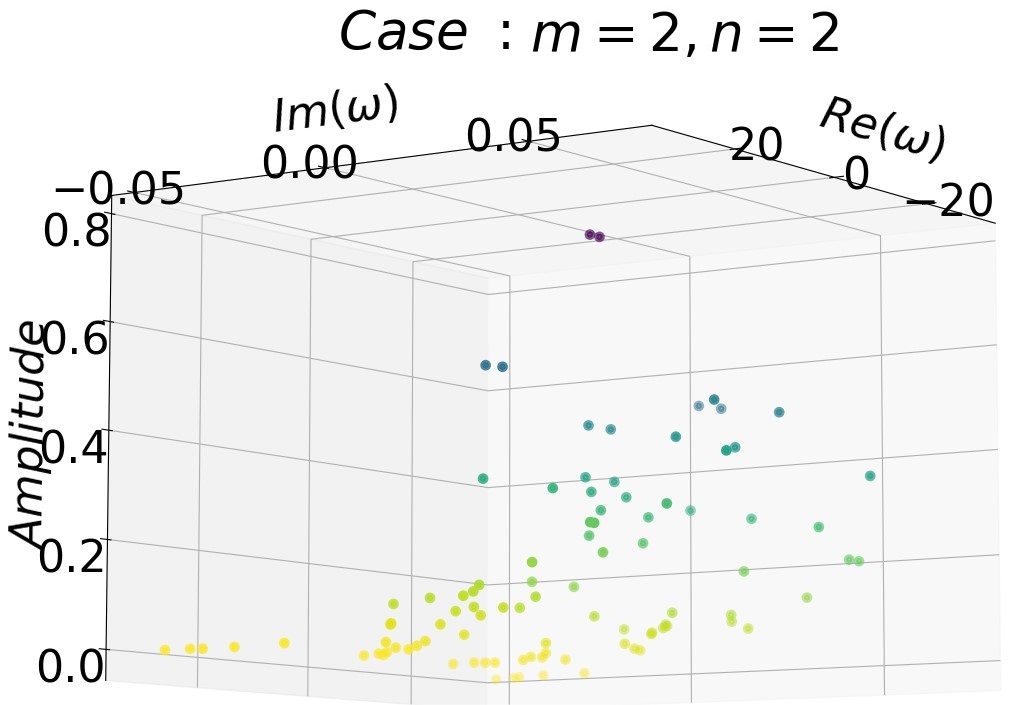} \hspace{.5cm}
    \includegraphics[width=0.45\textwidth]{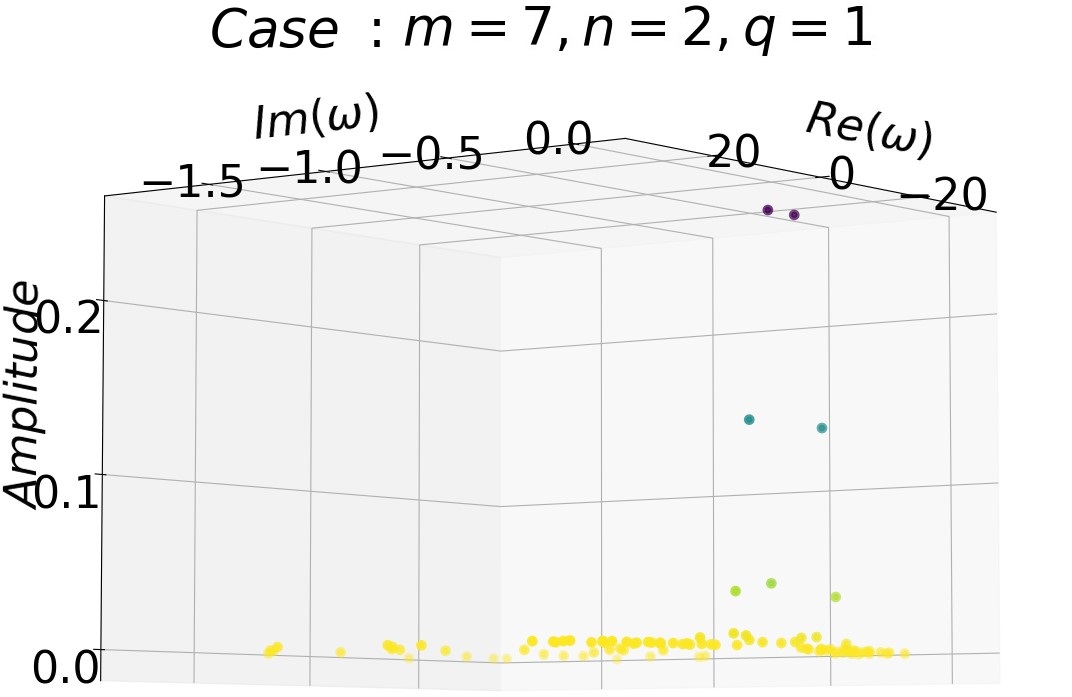}
    \caption{Amplitudes $A_n$ of fitting frequencies with dominant $\omega_{QNF}$ having the greatest value. }
    \label{fig:prony-amps}
\end{figure}

The Matrix (Prony) method returns a set of complex frequencies which are further analysed and sorted with respect to a magnitude similar to the Fourier technique. In Fig. \ref{fig:prony-amps}, we graphically present the amplitude of each mode obtained in matrix method (using Python). It is evident that only two frequencies (conjugate of each other) have the highest magnitude hence most dominating.
Thus the QNFs in their order of dominance can be identified. 
We shall not show the amplitude plots for any other cases for brevity.	

\subsubsection{Direct Integration}

In the Direct Integration method, the characteristic or the master differential equation Eq. (\ref{eq:Sch}) is numerically integrated using purely outgoing boundary conditions. This technique was first used by Chandrasekhar and S. Detweiler \cite{Chandrasekhar:1975zza} in 1975.
We essentially follow the steps described in \cite{Aneesh:2018hlp,DuttaRoy:2019hij}. 
As our potential is symmetric about the wormhole throat ($l=0$) (in both 4D and 5D cases) our solution can be of symmetric or antisymmetric kind. For symmetric (anti-symmetric) solution we should use $R'(0)=0$ ($R(0)=0$).
Note that, the asymptotic solution near $l \ra \infty$ can be expanded as
\begin{equation}
    R^+=e^{i\Omega l}\sum_{k=0}^{\infty}\frac{A^+_k}{l^k} ; ~~~~ \O^2 = \o^2 - q^2, ~~~~  \la{eq:DI-far}
\end{equation} 
which represent purely outgoing wave. 
However, near the throat or at some finite distance from the throat, expansion should contain both ingoing and outgoing waves, given by,
\begin{equation}
    R(l)=e^{i\Omega l}\sum_{k=0}^{\infty}\frac{A^+_k}{l^k} + e^{-i\Omega l}\sum_{k=0}^{\infty}\frac{A^-_k}{l^k}. \la{eq:DI-near}
\end{equation} 
By putting $R(l)$ in the field equation, one gets the following recurrence relations (at large $l_0^2 >> b_0^2$),
 \be
    A^{\pm}_{k+1}= \pm \frac{\{k(k+1) - m(m+1)\}A^{\pm}_{k} + (n-1)b_0^n A^{\pm}_{k-n}}{2i\O (k+1)} \la{eq:recur}
 \ee
This gives all the $A^{\pm}_{k}$ in terms of $A^{\pm}_{0}$.
After we integrate Eq. (\ref{eq:Sch}) from $l=0$ to $l=l_0$, we match the numerically found $R_{num}(l)$ and $R'_{num}(l)$ with Eq. (\ref{eq:DI-near}) and it's derivative at $l_0$. This leads to the following matching conditions 
\begin{equation}
    R_{num}(l_0)=e^{i\Omega l_0}\sum_{k=0}^{\infty}\frac{A^+_k}{l_0^k} + e^{-i\Omega l_0}\sum_{k=0}^{\infty}\frac{A^-_k}{l_0^k}, \la{eq:match-1}
\end{equation} 
\begin{equation}
    R'_{num}(l_0)=e^{i\Omega l_0}\sum_{k=0}^{\infty}\frac{A^+_k}{l_0^k} \le(i\O-\frac{k}{l_0}\ri) + e^{-i\Omega l_0}\sum_{k=0}^{\infty}\frac{A^-_k}{l_0^k} \le(-i\O-\frac{k}{l_0}\ri). \la{eq:match-2}
\end{equation} 
Eliminating $A^+_0$ from Eq. (\ref{eq:match-1}) and Eq. (\ref{eq:match-2}), we get an expression of $A^-_0$ as a function of $l_0$ and $\o$. Roots of the equation $A^-_0 = 0$, in the large $l_0$ limit, gives us the QNFs. 
The stability of the solutions is checked by verifying that varying $l_0$ does not considerably change the QNF values. We have only considered QNFs corresponding to the symmetric solutions, which have low damping.

%


%
%

\section{Results}\label{sec:Results}

Our focus will be $n=2$ or pure EB model as we are going to compare this results with 5D scenario. 
However, we will also address higher $n$-geometries (4D-GEB) briefly for completeness and to extend the results presented in \cite{DuttaRoy:2019hij}. 
\subsection{4D wormhole: varying $n$ and $m$}

\begin{figure}[h]
    \centering
    \includegraphics[width=.475\textwidth]{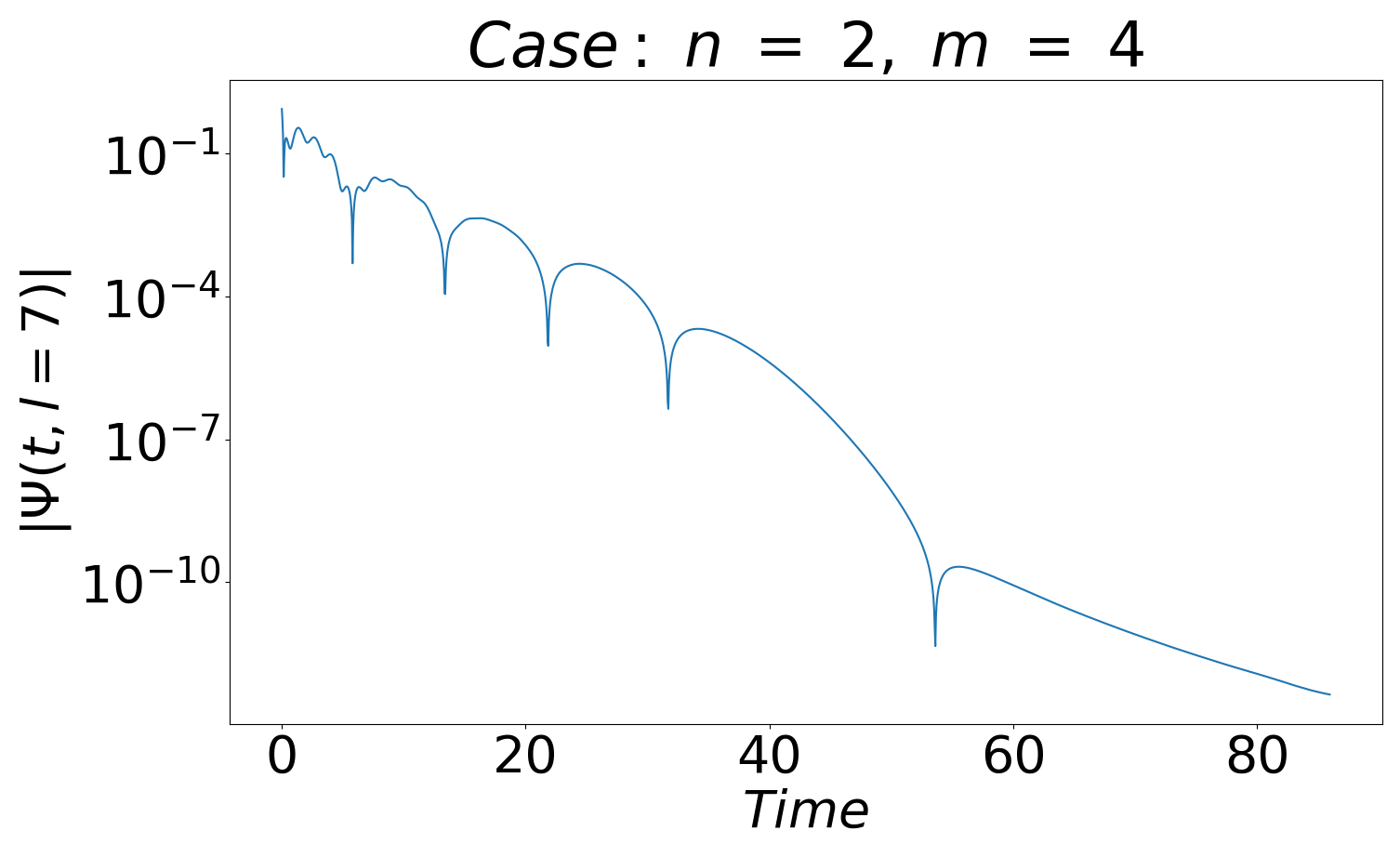} 
\hspace{.5cm}
   \includegraphics[width=.475\textwidth]{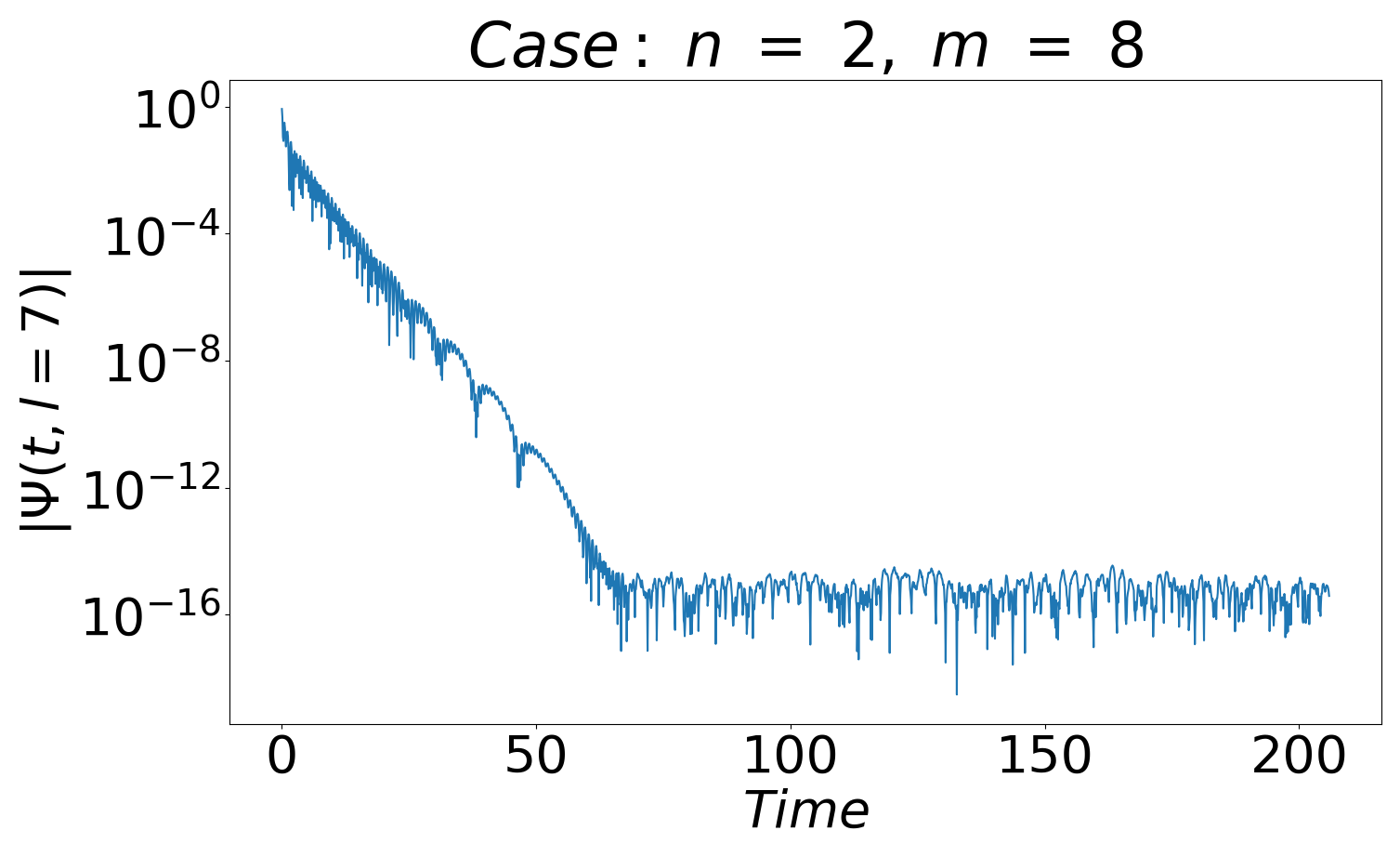} 
    \caption{Time domain Spectrum for $n=2, m=4$ and $n=2, m=8$.}
    \label{fig:n2}
\end{figure}

Fig. \ref{fig:n2} shows that for 4D-EB wormhole ($n=2$) the damped oscillatory region is less prominent for lower values of $m$. Therefore, QNF values extracted from these evolutions, using Prony fitting, are sensitive to the choice of beginning and end of QNM oscillation. 

\begin{figure}[h]
    \centering
    \includegraphics[width=.475\textwidth]{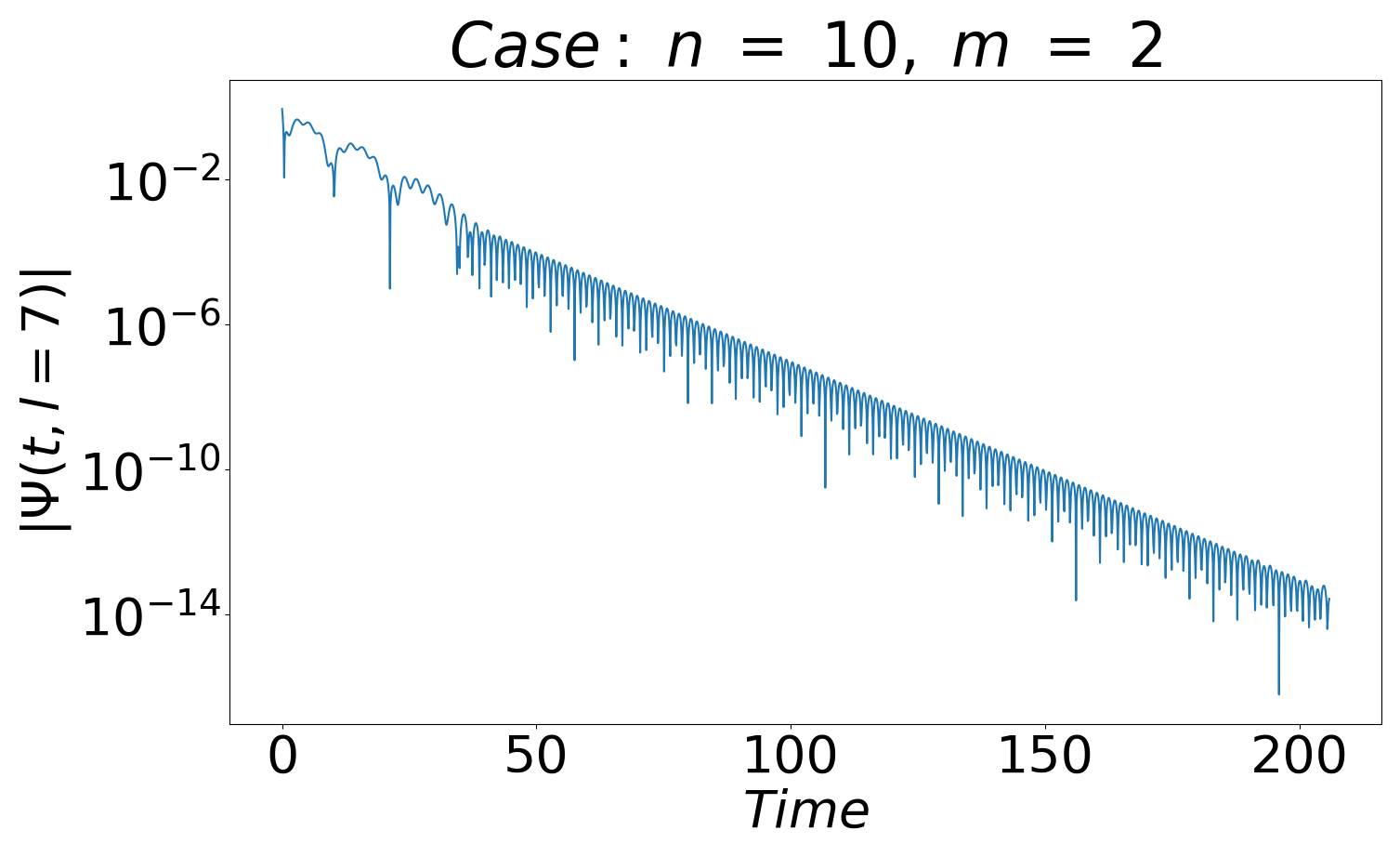} \hspace{.5cm}
    \includegraphics[width=.475\textwidth]{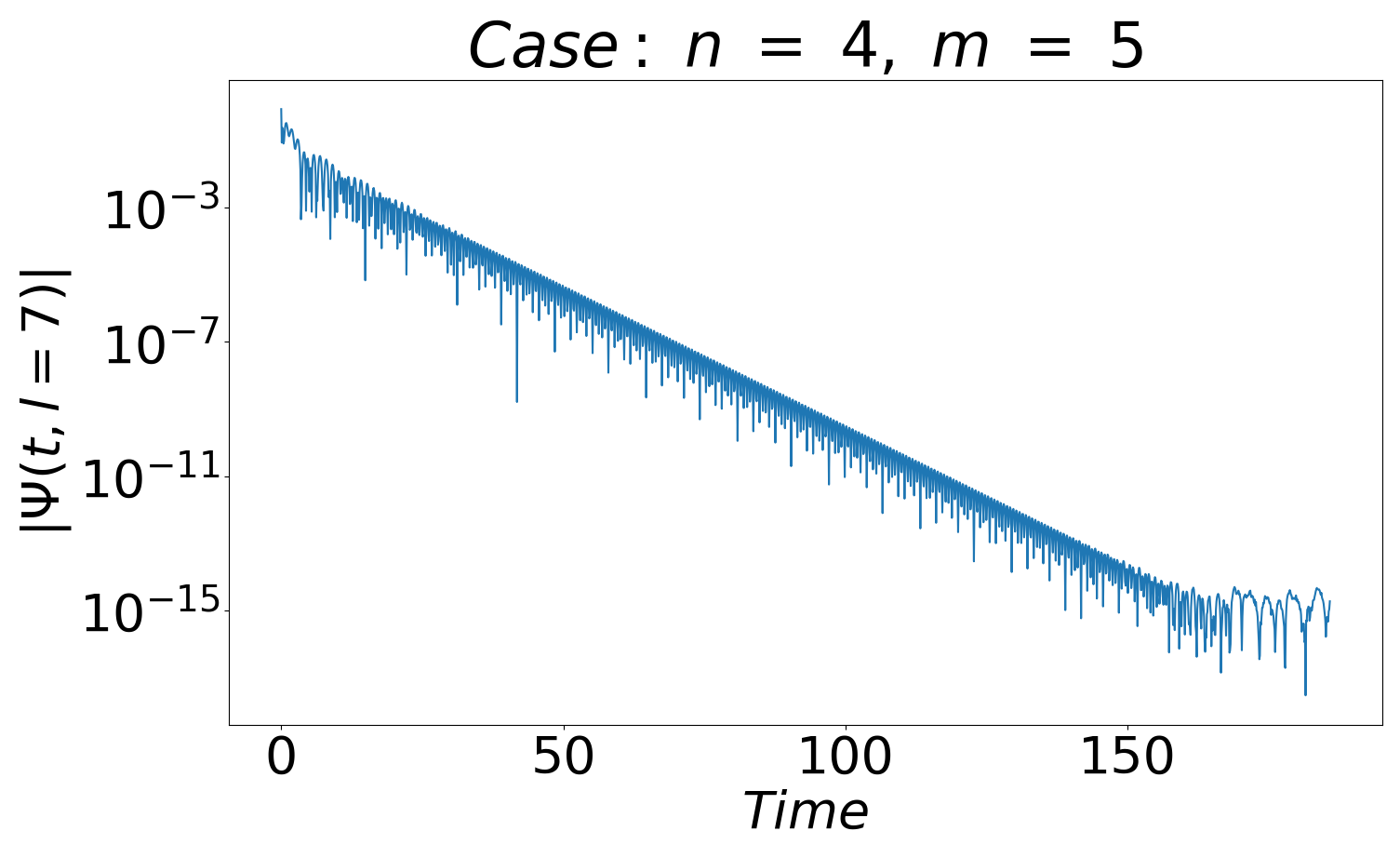}
    \caption{Time domain Spectrum for $n=10, m=2$ and $n=4, m=5$.}
    \label{fig:n10-4}
\end{figure}

Fig. \ref{fig:n10-4} shows time evoluton for a {\em steep-neck} 4D-GEB geometry with $n=10$ (for $m=2$) and $n=4$ (for $m=5$). 
Comparison with Fig. \ref{fig:Py-Mat} shows that, with increasing $n$, the beginning of the QN ringing domain has not changed much but the end is delayed considerably i.e. the tail appears much later for a higher value of $n$. Whereas with increasing $m$, the QNM oscillation gets triggered earlier. These particular features could be useful signature in detecting the shape of the GEB wormholes apart from those reported in \cite{DuttaRoy:2019hij}. 

\begin{figure}[h]
    \centering
    \includegraphics[width=.8\textwidth]{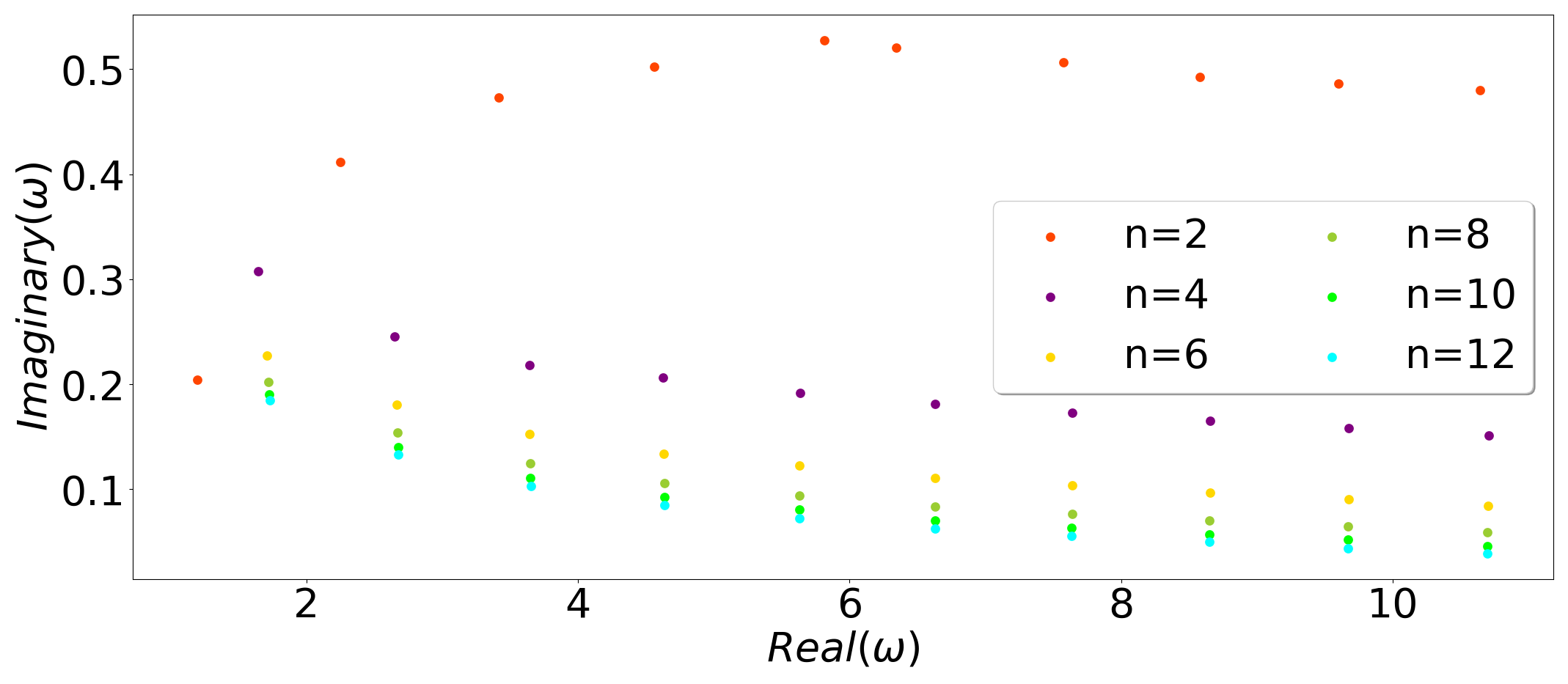} 
    \caption{Plot of $\omega_{QNF}$, real versus (magnitude of) imaginary part for different $n$ and $m$ values. From left to right, the $m$ value increases. }
    \label{fig:qnf_4dgeb}
\end{figure}

The (dominant) quasi-normal frequencies for various $m$ (angular momentum) and $n$ (steep-neck parameter) values in 4D-GEB have been plotted (real versus the absolute value of imaginary) in Fig. \ref{fig:qnf_4dgeb}. We clearly see that the features of $n=2$ is markedly different from $n>2$ scenario. Fig. \ref{fig:qnf_4dgeb} essentially reproduces results found in \cite{DuttaRoy:2019hij} and establishes the accuracy of our numerical computation. For a detailed discussion on Fig. \ref{fig:qnf_4dgeb}, we urge the reader to consult \cite{DuttaRoy:2019hij}.

\subsection{5D wormhole: $n=2$, varying $m$ and $q$}

\begin{figure}[h]
    \centering
    \includegraphics[width=.475\textwidth]{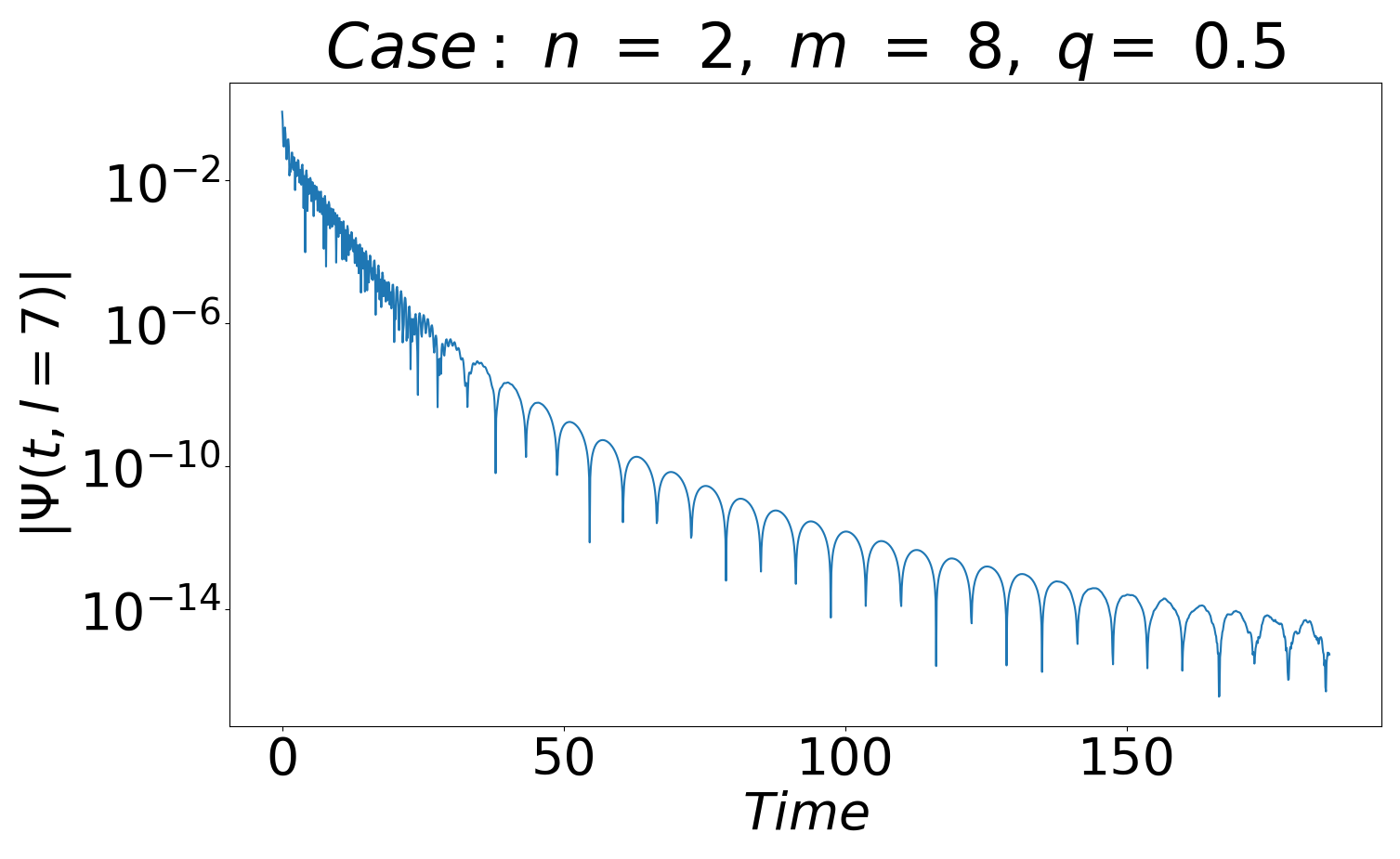}     
\hspace{.5cm}
    \includegraphics[width=.475\textwidth]{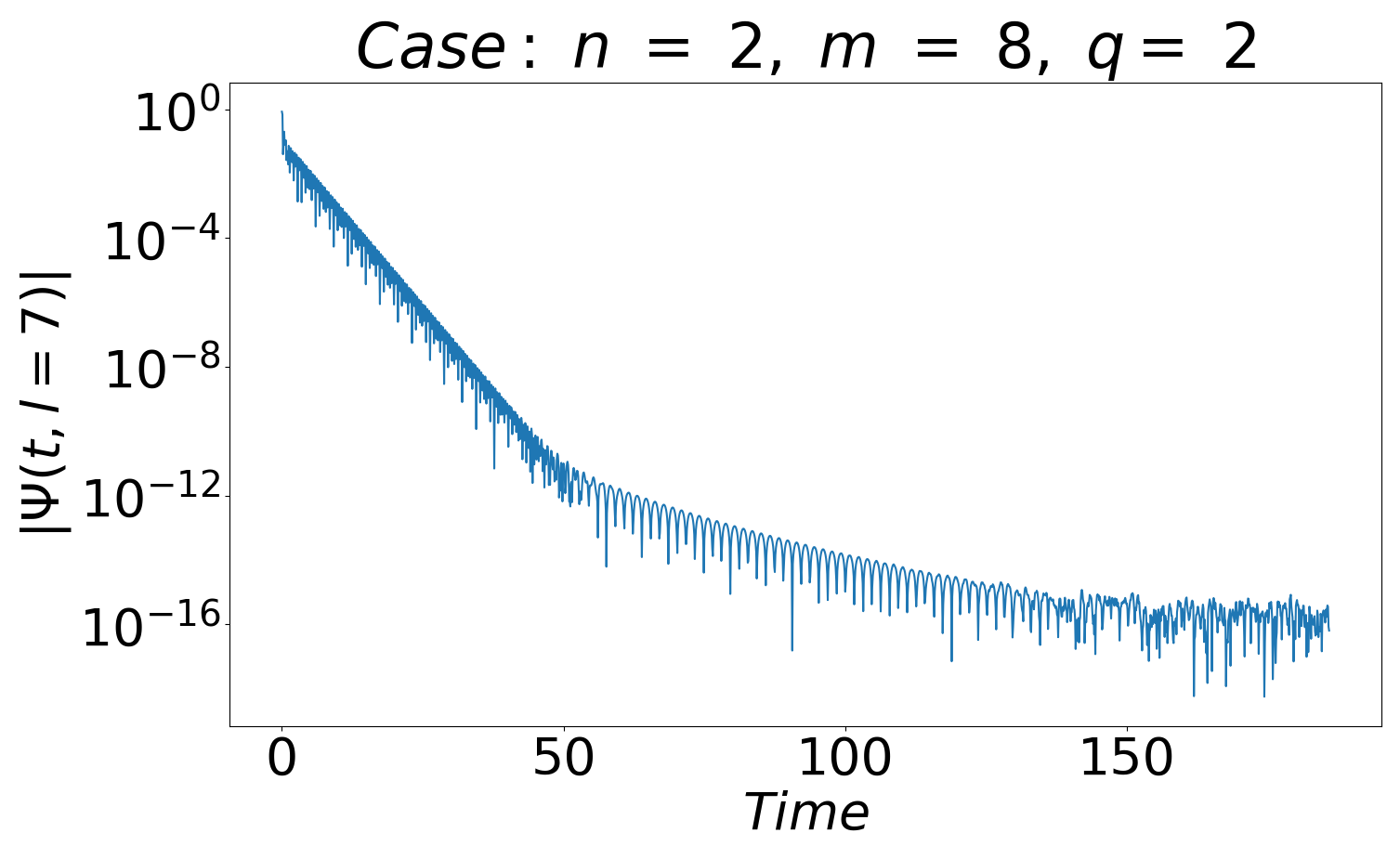} 
    \caption{Time domain spectrum for n=2, m=8 and q=0.5 (left), q=2 (right). }
    \label{fig:n2m8q}
\end{figure}

In the context of the 5D model, as argued earlier, we focus on the $n=2$ scenario. Incidentally, the effects of the extra dimension (i.e. of varying $q$-value or the  {\em massiveness} coming from the warped extra dimension) on the time evolution, is more striking for higher values of $m$. 
Let us look at the time domain profile for $n=2$, $m=8$ (any other value of $m$ will do) with varying $q$ values. Fig. \ref{fig:n2m8q}, shows the remarkable changes in the time evolution profile of the wave amplitude for four different values of $q$. 

\begin{figure}[h]
    \centering
    \includegraphics[width=.45\textwidth,height = 4cm]{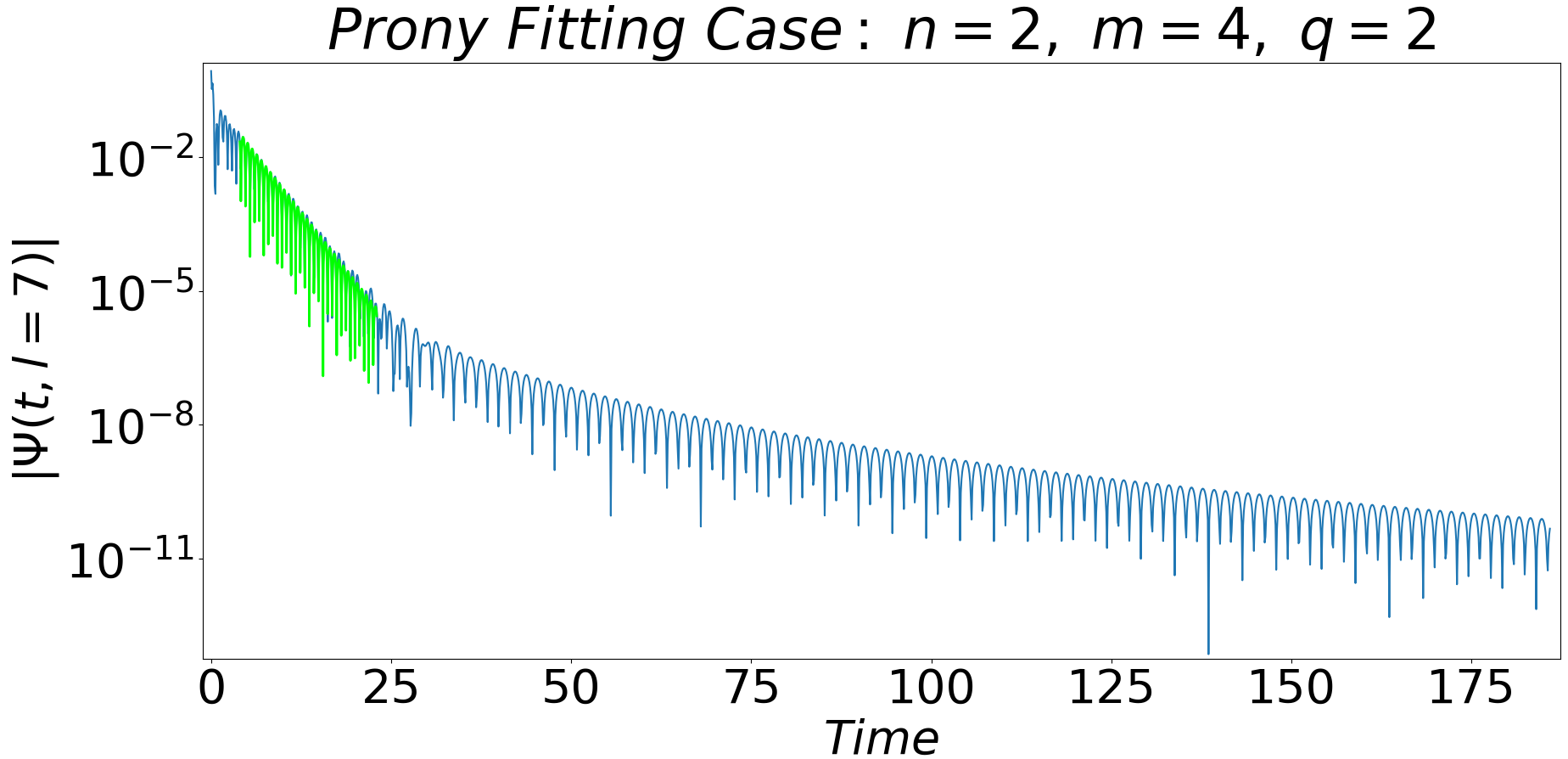} \hspace{.4cm}
    \includegraphics[width=.45\textwidth,height = 4cm]{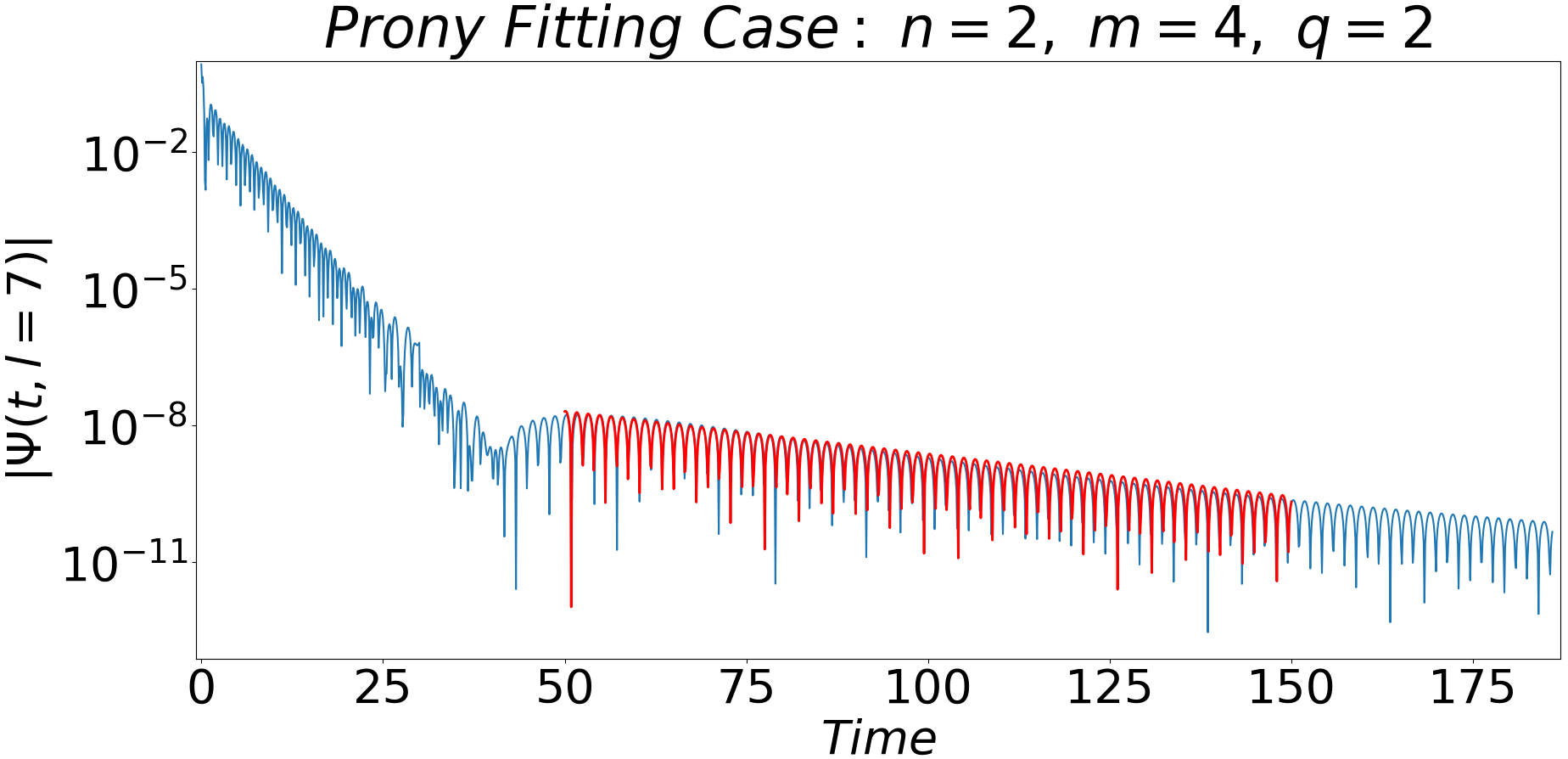} 
    \caption{Time domain spectrum and dominant QNM fit at early (left) and late (right) QNM era.}
    \label{fig:2qnf}
\end{figure}

Even for small (but non-zero) values of $q = 0.5$, the 4D behaviour (which is equivalent to setting $q=0$) is lost. This is expected as there is an interplay between $m$ and $q$.
Interestingly, for $q> b^{-1}_0$ (here we have taken $b_0=1$) , the semi-log plots clearly reveal that the QNM era is divided into  {\em two} parts (almost as if two linear regions with different slopes are joined at a kink) that are dominated by two different QNF modes. 
Notably, in the latter region, the most dominant QNM is characterised by  $Re(\o) \sim q$ with a small imaginary part as depicted in the tables given below. 
As time evolves, eventually, when the early dominant modes decay, the late-QNM emerges.
To further reveal the late-QNM region, we present Fig. \ref{fig:2qnf}. 
Here, in the left plot we show a perfect fit of the early-QNM region using the dominant QNF found to be $\o^{(E)} = 5.335 + i 0.41$. In the right plot, we have fitted the wave amplitude, after subtracting that dominant early-QNM, with the dominant late-QNM given by $\o^{(L)} = 2.0078 + i 0.0043$.
The order of their dominance has also been confirmed from their amplitudes using matrix Prony method as mentioned earlier. 
The (almost) purely real frequencies, in the late QNM era, are similar to the quasi-resonances found in  \cite{Konoplya:2004wg,Ohashi:2004wr}. 
Though imaginary part is non-zero but very small for the QNMs we report here, it is easy to see that (from Table \ref{tab:n2-early} and \ref{tab:n2-late}), for larger and larger $q$, the imaginary part indeed tends to zero and we have exact quasi-resonances.
Note that, it's not the existence of the quasi-resonances, but the existence of two QNM era, where the {\em late} QNM region is dominated by the almost quasi-resonant QNM, that we emphasize on as signature of the warped extra dimension in the ringing of a effective 4D-EB wormhole.

In \autoref{tab:n2-early} and \autoref{tab:n2-late}, we tabulated early and late dominant QNFs respectively, for $n=2$, various $q$-momentum and angular momenta $m$, determined using different methods. 
For brevity, we have denoted the early dominant QNF as $\o^{(E)}$ and the late dominant QNF as $\o^{(L)}$.
Note that WKB method only matches with early QNMs.

\begin{table}[h]
    \centering
    \caption{Early time dominant QNF-- $\omega^{(E)}$ values for various modes at n=2.}
    \begin{tabular}{ |  m{.5cm} | m{.5cm} | m{4cm}| m{4cm} | m{4cm} |} 
\hline
 \textbf{m} & \textbf{q}  & \textbf{Prony } &  \textbf{Direct Integration }& \textbf{WKB } \\ 
\hline
\hline
 \multirow{5}{*}{1}	 & 	0 & 0.8 32130 -i0.226690  & 1.624584 -i0.219350 & 1.617350 - i0.250251\\ 
\cline{2-5}
 	 & 	0.5 &  0.750281 -i0.188820 & 1.851582 -i0.188352 & 1.879640 - i0.242017\\  
\cline{2-5}
 	 & 	1 & 1.335939 -i0.215243 & 2.053928 -i0.215267 & 2.058170 - i0.185868\\  
\cline{2-5}
	 & 	2 &  2.621701 -i0.123791 & 2.621708 -i0.123782 & 2.672090 - i0.124239\\  
\cline{2-5}
 	 & 	5 & 5.396131 -i0.036641  & 5.396137 -i0.036642 & 5.394870 -i0.038862\\  
\hline
\multirow{5}{*}{2}	 & 	0 &  2.250079 -i0.457395  & 2.712579 -i0.445679 & 2.629721 - i0.424325\\ 
\cline{2-5}
 	 & 	0.5 &  1.505729 -i0.317217 & 2.810257 -i0.312587 & 2.741552 - i0.315845\\  
\cline{2-5}
 	 & 	1 & 2.528403 -i0.279339 & 2.948405 -i0.279358 & 2.871002 - i0.282586\\  
\cline{2-5}
	 & 	2 &  3.195762 -i0.231845 & 3.194526 -i0.221053 & 3.343490 - i0.222976\\  
\cline{2-5}
 	 & 	5 & 6.089024 -i0.304504  & 6.129243 -i0.315208 & 6.171861 - i0.303122\\  
\hline
 \multirow{5}{*}{5}	 & 	0 & 5.590286 -i0.527592  & 5.582526 -i0.512567 & 5.590768 - i0.505916\\ 
\cline{2-5}
	 & 	0.5 & 5.608272 -i0.515912 & 5.611273 -i0.518386  & 5.612845 - i0.503921\\  
\cline{2-5}
	 & 	1 &  5.732418 -i0.514532 & 5.727413 -i0.503953 & 5.678741 - i0.498073 \\  
\cline{2-5}
 	 & 	2 & 5.913451 -i0.477296  & 5.922475 -i0.476795  & 5.935240 - i0.476548\\  
\cline{2-5}
 	 & 	5 & 7.480281 -i0.367137&  7.480376 -i0.369172  & 7.492832 - i0.377485 \\  
\hline
 \multirow{5}{*}{8}	 & 	0 & 8.579942 -i0.492730 & 8.529875 -i0.491728 & 8.558772 - i0.492544 \\ 
\cline{2-5}
 	 & 	0.5 & 8.621930 -i0.486822 &  8.624282 -i0.482853 & 8.573310 - i0.491692 \\  
\cline{2-5}
	 & 	1 & 8.643952 -i0.481875 &   8.647258 -i0.484326  & 8.616820 - i0.489164\\  
\cline{2-5}
	 & 	2 & 8.979098 -i0.468332 &   8.979096 -i0.478331  & 8.788610 - i0.469402\\  
\cline{2-5}
	 & 	5 & 9.979701 -i0.312355&  9.978305 -i0.327316 & 9.90901 - i0.324066 \\  
\hline
 \end{tabular}
    
    \label{tab:n2-early}    
\end{table}


\begin{table}[h]
    \centering
    \caption{Late time dominant QNF-- $\omega^{(L)}$ values for various modes at n=2.}
    \begin{tabular}{ |  m{.5cm} | m{.5cm} | m{4cm}| m{4cm}| } 
\hline
 \textbf{m} & \textbf{q}  & \textbf{Prony } &  \textbf{Direct Integration}\\ 
\hline
\hline

\multirow{5}{*}{1}	
& 	0.5  & 0.507324 -i0.030946 &0.507425 -i0.030862\\  
\cline{2-4}
 	 & 	1  & 1.001504 -i0.017169 & 1.001248 -i0.017066 \\  
\cline{2-4}
 	 & 	2  & 2.008260 -i0.003081 &2.008267 -i0.003019 \\  
\cline{2-4}
 & 	5  & 5.030260  -i0.000004 &5.030188 -i0.000009\\  
\hline
\multirow{5}{*}{2}	
& 	0.5  & 0.513874 -i0.055556 &0.512767 -i0.055676\\  
\cline{2-4}
 	 & 	1  & 1.004001 -i0.014508 & 1.005080 -i0.014216 \\  
\cline{2-4}
 	 & 	2  & 2.009095 -i0.004750 &2.008207 -i0.004302 \\  
\cline{2-4}
 & 	5  & 5.036428  -i0.000003 &5.035281 -i0.000002\\  
\hline
\multirow{5}{*}{5}	
& 	0.5 &  0.525911  -i0.122378 &  0.525908  -i0.122350\\  
\cline{2-4}
 & 	1 & 1.016924 -i0.037482 &  1.016979 -i0.039472\\  
\cline{2-4}
 & 	2 & 2.006454 -i0.008466 & 2.006424 -i0.008492\\  
\cline{2-4}
 & 	5 &  5.004510 -i0.000001 &   5.004522 -i0.000002\\  
\hline
 \multirow{5}{*}{8}	
& 	0.5 & 0.504295  -i0.156266   & 0.504383  -i0.156142\\  
\cline{2-4}
 & 	1 & 1.032865 -i0.036472 & 1.030675 -i0.036488\\  
\cline{2-4}
 & 	2   & 2.001427  -i0.001418  & 2.001445  -i0.001433\\  
\cline{2-4}
 & 	5 &  5.004831  -i0.000000  & 5.004853  -i0.000000 \\  
\hline
 \end{tabular}
    
    \label{tab:n2-late}
\end{table}

\begin{figure}
    \centering
    \includegraphics[width=0.5\textwidth, height = 4 cm]{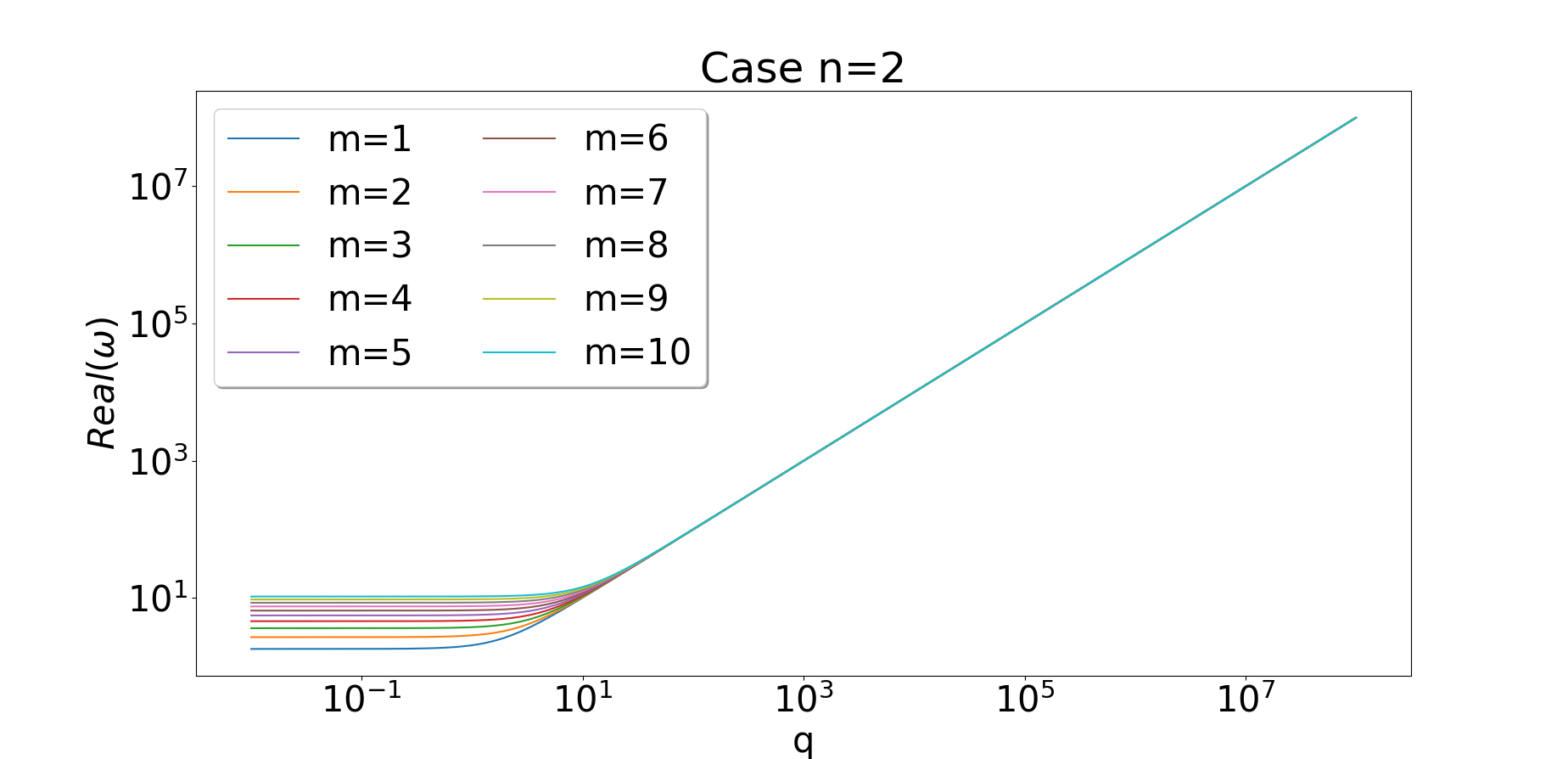}~~~~
    \includegraphics[width=0.45\textwidth, height = 4 cm]{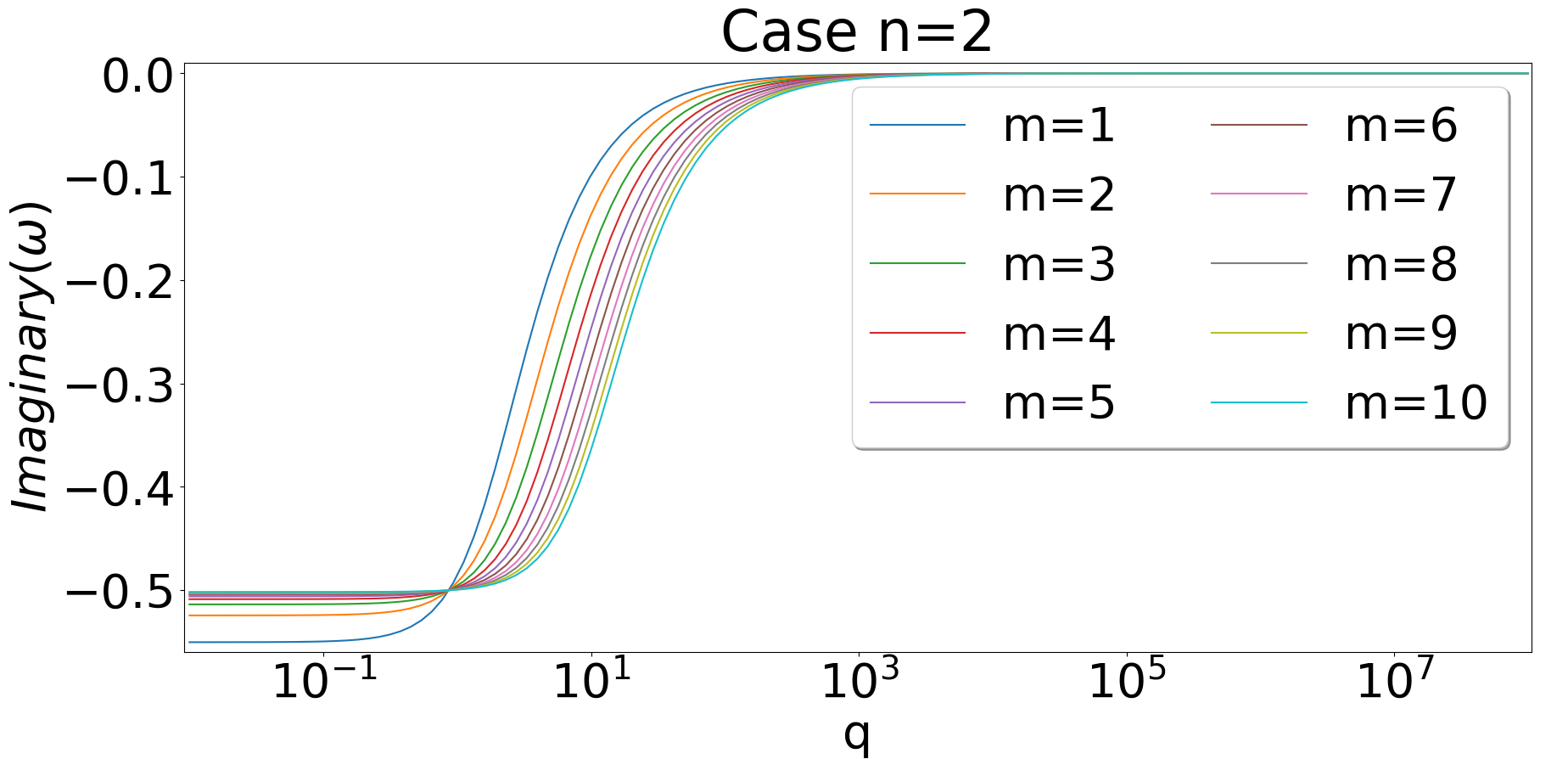} 
    \caption{Plot of $Re(QNF)$ and $Im(QNF)$ for $n=2$ with varying $q$ and $m$.  }
    \label{fig:W-q}
\end{figure}

In \autoref{tab:n2-early}, the QNF values for $q=0$ do match upto three digits after the decimal point with the 4D-GEB QNF values reported in \cite{DuttaRoy:2019hij}. 
Thus proving the accuracy of our numerical computation. 
The Prony method fails to provide accurate determination of early QNFs for low $m$ values because of small duration.
However, using larger numerical value for $b_0$, duration increases and prony method gives better results for low $m$ values as well.
Also, as $q$ increases, accuracy improves.
For non-zero $q$, the dominant QNF in the early QNM era gets larger with increasing $q$.  
Table \ref{tab:n2-late} shows that as the $q$ value increases, the imaginary part of the fundamental mode tends to zero asymptotically while the real part approaches $q$.
Apparently, from Eq. (\ref{eq:5D-pot}), this behaviour is expected if one takes the $q>>m$ limit.
Fig. \ref{fig:W-q} shows how the real and the imaginary part of the QNF varies with varying $q$.

Note that in  \cite{Konoplya:2004wg}, it was found that, for massive scalar fields in the black hole background, quasi resonance is achieved above certain threshold value of the mass whereas we are getting the similar behaviour in the $q \ra \infty$ limit. 
Further, the short lived early QNM era was also not reported there.
It is crucial to identify the duration of the QNM era to determine accurate values of QNF. 
Note that for low angular momentum values $m$, the duration of the early QNM era is small thus difficult to detect.
Thus one might guess that even in case of massive scalar field there exists an early QNM era.
However, we did look at the time domain profiles in detail for massive perturbation of black holes but did not find such behaviour.
Coming back to our model, for $q < b^{-1}$, duration of early QNM decreases indefinitely with decreasing $q$. 
For such low values of $q$, the timespan of the dominant QNF in the the early QNM era becomes smaller than our algorithm's precision limit in the Prony method. This limitation does not show up for the other methods so they generate efficient values for QNF in those cases.


\section{Discussion}\label{sec:Dis}

Ellis-Bronnikov wormhole (and it's generalised versions) embedded in warped braneworld background has been shown to be supported by positive energy density matter in presence of a decaying warp factor. 
The violation of the weak energy condition can be minimised arbitrarily. 
Earlier we have studied particle trajectories and geodesic congruences in such spacetimes.
The recent observations suggest that one way to understand true nature of the ultra-compact objects is through their quasi-normal ringing. This method could potentially distinguish among black holes and possible black hole mimickers such as wormholes.
Here we analyse the QNMs of the 5D-WEB wormhole spacetime while looking for distinguishing features of the warped extra dimension and the wormole parameter.
The work done and the results found that reveal the effects of the warped extra dimension (through effective mass $q$) and the wormhole (steep-neck) parameter $n$ on the time domain profile and the QNFs. We summarise the key findings below in a systematic manner.
 %
\begin{itemize}
    \item The nature of the effective potential is almost similar in both four and five dimensions with a crucial difference that in 5D, the potential does not vanish asymptotically. 
    \item The momentum eigenvalue along the fifth dimension $q$, projected on the 4D geometry acts as an effective mass. We solved the corresponding eigenvalue problem and found that $q$ takes non-negative continuous values.
    \item Assuming suitable values of $q$, we then determined the QNFs analytically using WKB formula and numerically using the Prony method and the direct integration method. The results for 4D-GEB model match to 3rd decimal order with earlier reports. We have used both Python and Matlab for numerical computation of QNFs.    
        \item For 4D-GEB spacetimes, the time domain profile has three prominent regions (the initial portion, the QNM era  and an asymptotic  tail. Apart from the observations made in \cite{DuttaRoy:2019hij}, we notice that the QNM ringing appears earlier for higher angular momentum $m$ and the tail appears later for higher values of $n$. 
    \item Remarkably, the time domain profile changes considerably in 5D-WEB scenario. 
 The QNM era is divided into two parts with two different dominant QNFs. 
The real part of the `early QNM' (for fixed $m$) increases with increasing $q$ value whereas the real part of the `late QNM' is close to $q$ value.
Also, the dampening (decided by the imaginary part of QNF) of the late QNM is much slower than that of early QNM. 
\item With increasing $q$, the late QNM eventually becomes arbitrarily long lived. This arbitrarily long lived modes emerges once the dominant (early) QNM decays away. These so-called quasi-resonances were observed earlier for massive fields in black hole backgrounds
. 
Further, the tail appears much later compared to the 4D scenario.

\end{itemize}

Note that, the novel feature that could be seen in the time domain profile of the perturbation is not of quantitative or parametric type. There are {\em two QNM eras} observed, instead of one, whose durations naturally vary for varying $q$. We believe such behaviour is predicted or reported in the literature for the first time.
Hence, one may conclude that this feature of two different dominating QNFs in the ringing profile, if observed, could provide indirect evidence of existence (though not conclusive) of wormhole as well as five-dimensional warped geometry. 
It is always difficult to build physical intuition about the nature of the QNMs once they are determined numerically.
However we have seen that it's the interplay between $m$ and $q$ that is responsible for the splitting of the QNM era.
Further, one might find similar splitting in other equivalent and preferably more generic scenarios.
We look forward to report on this in future.

\section*{Acknowledgement}

We thank Dr. Poulami Dutta Roy and Prof. Sayan Kar for useful discussions and correspondence.
SG thanks BIT Mesra for financial assistance through seed money scheme.

\section*{Bibliography} 

\bibliographystyle{unsrt}
\bibliography{References}


\end{document}